\documentclass[10pt,letterpaper]{article}
\usepackage{color}
\usepackage{multicol}
\usepackage[letterpaper, margin=1in]{geometry}
\usepackage{subcaption}
\usepackage[overload]{empheq}

\usepackage{xcolor}

\usepackage{listings}
\usepackage{amsmath, amsthm, amssymb, wasysym, verbatim, bbm, graphics}
\usepackage{enumerate}
\usepackage{url}
\usepackage{mathtools}
\usepackage[colorlinks,linkcolor=blue]{hyperref}
\usepackage{xfrac}
\usepackage{float}
\usepackage{amsmath,bm}
\usepackage[title]{appendix}
\usepackage{todonotes}
\usepackage{subcaption}
\usepackage{multirow}
\usepackage{caption}
\usepackage{cleveref}
\usepackage{enumitem}
\usepackage{booktabs}
\usepackage{graphicx}
\usepackage{algorithm}
\usepackage{algpseudocode}
\usepackage[utf8]{inputenc}      % if you haven’t already
\DeclareUnicodeCharacter{0300}{\`{}}  % map U+0300 → \`

\graphicspath{{../figures/}}

\DeclareMathOperator*{\argmax}{arg\,max}

\usepackage[backend=biber,maxnames=4]{biblatex}
\newtheorem*{maintheorem*}{Main Theorem}
\theoremstyle{definition}{}

\allowdisplaybreaks

\numberwithin{equation}{section}

\usepackage[backend=biber,maxnames=4]{biblatex}
\begin{document}

\title{Kinetic Optimization of Magnetic Mirror Confinement: Beyond Classical Loss-Cone Theory}
\author{Lukas Einkemmer\thanks{Department of Mathematics, Universit\"at Innsbruck, Innsbruck, Austria} \and Martin Guerra\thanks{Department of Mathematics, University of Wisconsin-Madison, Madison, WI, USA} \and Qin Li\footnotemark[2] \and Leonardo Zepeda-N\'u\~nez\thanks{Google Research, Mountain View, CA, USA}}
\date{}

%\headers{Plasma Confinement for Magnetic Mirrors}{L. Einkemmer, M. Guerra, Q. Li, L. Zepeda-N\'u\~nez}

\maketitle

\begin{abstract}
Magnetic mirrors are among the conceptually simplest plasma confinement configurations and remain promising candidates for thermonuclear fusion. Their design requires shaping an externally applied magnetic field to confine plasma within an open-ended cylindrical device. In contrast to toroidally closed devices such as tokamaks and stellarators, confinement in magnetic mirrors depends intrinsically on kinetic mechanisms, particularly velocity-space trapping and particle loss through the open ends.

We formulate magnetic mirror design as a PDE-constrained optimization problem governed by a reduced multispecies drift-kinetic–Poisson model. The resulting optimization reveals two physical effects not captured by the classical loss-cone argument. First, the self-consistent electric field generated through Poisson coupling acts as a secondary confinement barrier and substantially alters particle retention in the nonlinear regime. Second, the optimized magnetic-field configuration depends qualitatively on the underlying kinetic model: an electron-only model favors an unconventional centrally peaked field, whereas the fully coupled electron–ion model recovers the classical boundary-peaked mirror configuration. These results demonstrate that optimal magnetic mirror design cannot be determined solely from loss-cone considerations, but must account for the self-consistent nonlinear kinetic dynamics of the plasma.
\end{abstract}

% ==========================================
% 1. INTRODUCTION
% ==========================================

\section{Introduction}
\label{sec:intro}
The confinement of high-temperature plasmas is a fundamental challenge in the development of controlled thermonuclear fusion~\cite{chen1984introduction}. The goal is to design a device capable of confining hot plasma particles at sufficiently high density and for sufficiently long times to sustain nuclear fusion reactions. While much of the research effort has focused on the design and control of tokamaks and stellarators, alternative designs have also shown significant promise. Among them are magnetic mirror devices~\cite{post1987magnetic}, which have recently attracted renewed attention and are undergoing active development~\cite{Realta}.

In contrast with relying on toroidal closure for confinement (as is done in tokamaks and stellarators), a magnetic mirror typically consists of a cylindrical device with a magnetic field strength that increases towards the ends. Such stronger fields exert a force on particles so the particles that travel towards the ends eventually reverse directions. It effectively generates two mirrors at the ends that bounce particles back and forth in the cylindrical chamber, thus confining particles.

Although conceptually simple, pure magnetic mirror devices are inherently lossy. Plasma particles {with insufficient velocity perpendicular to the magnetic field} can escape through the ends of the device. A central objective in magnetic mirror design is therefore to engineer magnetic field configurations that minimize particle leakage, thus maximizing long-time confinement. Mathematically, this objective naturally leads to an inverse design problem in which one seeks an optimal magnetic field configuration $\mathbf{B}$ maximizing plasma confinement:
\begin{equation}
\label{eq:objective_intro}
    \argmax_{\mathbf{B}} \mathcal{J}(\mathbf{B}),
\end{equation}
where $\mathcal{J}$ measures the amount of plasma retained inside the confinement region over a prescribed time horizon. {The classic concept of the ``loss cone'' states that particles that have an unfavorable ratio of the velocity parallel versus the velocity perpendicular to the magnetic field can escape, forming a depleted cone in phase space \cite{Kolmes2024loss_cone}.} Such theories are primarily based on single-particle trajectory arguments and largely neglect self-consistent collective effects generated by the plasma itself. In reality, however, the plasma system is intrinsically nonlinear: the collective charge distribution generates a self-consistent electric field, which in turn significantly modifies particle confinement and escape dynamics. Understanding the role of this self-consistent electrostatic effect in magnetic mirror optimization is the primary objective of this work.

The collective dynamics is described by Vlasov-type kinetic equations. While the most fundamental formulation is a multi-species $3\text{D}3\text{V}$ Vlasov system, the presence of a strong background magnetic field permits a substantial reduction of complexity. {Averaging over the fast circular motion of the particles perpendicular to the magnetic field (the gyromotion) results in a drift-kinetic equation that only depends on the velocity parallel to the magnetic field and the magnitude of the velocity perpendicular to it. The latter is expressed in terms of the magnetic moment $\mu$, and is an adiabatic invariant. It only enters the drift-kinetic equations as a (continuous) parameter. If we further assume a homogeneous plasma in the direction perpendicular to the magnetic field}, we obtain a $1\text{D}1\text{V}$ kinetic model parameterized by continuous $\mu$-slices~\cite{cary2009guiding,tyushev2025,JIMENEZ2024magnetic,hazeltine2003plasmaconfinement,sonnendrucker1999semi}. This reduced formulation preserves the essential kinetic features while significantly lowering the computational cost. Coupling this reduced model with the confinement objective~\eqref{eq:objective_intro} yields a PDE-constrained optimization framework for magnetic mirror design, which we will investigate in this work.

Our findings reveal several nonclassical confinement phenomena. First, confinement performance cannot be accurately predicted solely through classical mirror-ratio heuristics: magnetic field configurations with similar mirror ratios may exhibit substantially different confinement behavior once self-consistent electrostatic effects are incorporated. In particular, the induced electric potential forms an additional confinement barrier that significantly alters particle escape dynamics ({especially for electrons}). Second, through running PDE-constrained optimization using reverse-mode automatic differentiation, we discover that the optimal magnetic topology depends strongly on the plasma composition. In the single-species regime, optimized configuration turns out to favor nonclassical centrally-peaked structures that have a double-well form. This deviates substantially from traditional design of single-well mirror profiles. In the multi-species setting, the disparate electron-ion mass ratio introduces strongly separated confinement timescales, leading to a two-stage leakage process and the recovery of the classical one-well design.

The remainder of the paper is organized as follows. In Section~\ref{sec:math_formulation}, we derive the reduced kinetic formulation and present the PDE-constrained optimization framework. In Section~\ref{sec:loss_cone}, we investigate the emergence of nonlinear self-consistent confinement effects and we demonstrate the limitations of classical loss-cone predictions through extensive numerical experiments. Section~\ref{sec:numerics} presents optimized magnetic confinement configurations for both single-species and multi-species plasmas.

\subsection{Related Work}
\label{sec:related_work}
The mathematical formulation and numerical simulation of magnetic mirror plasmas intersect several active areas of research, including reduced kinetic modeling, structure-preserving numerical methods for Vlasov equations, and PDE-constrained optimization through differentiable simulation.

The foundational physics of magnetic mirrors and the associated loss-cone mechanism have long been studied theoretically~\cite{post1987magnetic,Pastukhov1974,brizard2007foundations}. Given that the charged particles undergo rapid gyromotion around magnetic field lines, direct simulation of the full multi-species $3\text{D}3\text{V}$ Vlasov system is computationally demanding. Modern multiscale plasma modeling therefore heavily exploits asymptotic reductions based on strong background magnetic fields. In particular, by averaging over the fast cyclotron motion and exploiting the adiabatic invariance of the magnetic moment, one obtains drift-kinetic and gyrokinetic descriptions that substantially reduce computational complexity while preserving the essential confinement dynamics~\cite{cary2009guiding,brizard2007foundations}.
This formulation follows a similar philosophy by adopting a continuous $\mu$-parameterized reduced kinetic description that isolates the slow confinement dynamics while averaging over fast gyromotion.

The reduced multi-species Vlasov--Poisson system nevertheless remains numerically challenging due to the strong disparity between electron and ion time scales. Standard explicit Eulerian discretizations are severely restricted by the Courant--Friedrichs--Lewy (CFL) condition associated with the fastest characteristic speeds~\cite{filbet2001conservative}. Consequently, Semi-Lagrangian (SL) methods have become a central tool in kinetic plasma simulations~\cite{sonnendrucker1999semi,crouseilles2010conservative,CROUSEILLES2026SAV,EM2024,EM2025}. By transporting the distribution function along characteristic trajectories, SL schemes substantially relax the CFL constraint while retaining favorable conservation properties and positivity when combined with suitable interpolation procedures.

In parallel, plasma confinement optimization has increasingly benefited from advances in PDE-constrained optimization, scientific machine learning, and differentiable programming~\cite{nature2022,Einkemmer2024,Einkemmer2024beam,Einkemmer2025,albi2025instantaneous,crouseilles2025control,Bartsch_KRM,albi2025robust,lu2025dynamical}. Traditional approaches to magnetic topology optimization, including the design of quasi-isodynamic stellarators and mirror configurations, often rely on surrogate models or gradient-free heuristics due to the large dimensionality and complexity of the design space~\cite{wei2026lowdimensional,albert2020}. More recently, differentiable simulation frameworks have enabled gradient-based optimization directly through discretized plasma solvers by applying reverse-mode automatic differentiation to the numerical pipeline~\cite{guerra2026metric,joglekar2026differentiable,Bartsch_Knopf,Lovbak_Samaey,LiWangYang_MC}. Such approaches permit the computation of discrete gradients of macroscopic quantities, including confinement metrics and energy functionals, with respect to control parameters and boundary conditions.

While reduced kinetic modeling, structure-preserving Vlasov solvers, and differentiable PDE optimization have each been extensively investigated independently, our goal here is to study the effect of kinetic nonlinearities on the optimal magnetic field configuration. In particular, the role of the self-consistent electric field in reshaping confinement behavior and how it modifies loss cone dynamics is investigated. Our code enables a complete end-to-end differentiable optimization of magnetic mirrors within a three-dimensional drift-kinetic model.

% ==========================================
% 2. MATHEMATICAL FORMULATION
% ==========================================

\section{Mathematical Formulation}
\label{sec:math_formulation}

In this section, we derive a reduced kinetic formulation for magnetic mirror confinement based on the adiabatic invariance of the magnetic moment. Starting from the full particle dynamics under a strong magnetic field, we perform a guiding-center reduction that leads to a $1\text{D}1\text{V}$ Vlasov--Poisson system parameterized by the magnetic moment $\mu$. We then formulate the corresponding PDE-constrained optimization problem for magnetic mirror design.

\subsection{Guiding-center reduction and magnetic moment}

We begin with the dynamics of a single charged particle under an external magnetic field $\mathbf{B}$. Writing
\[
\mathbf{B} = B_0 \mathbf{b},
\]
where $B_0 = |\mathbf{B}|$ denotes the magnetic field strength and $\mathbf{b}$ is the associated unit direction vector. The particle velocity can then be decomposed into parallel and perpendicular components:
\[
\mathbf{v} = v_{\parallel}\mathbf{b} + \mathbf{v}_{\perp},
\qquad
v_{\parallel} = \mathbf{v}\cdot \mathbf{b},
\qquad
|\mathbf{v}_{\perp}| = v_{\perp}.
\]
The corresponding magnetic moment is defined by
\[
\mu = \frac{v_{\perp}^{2}}{2B_0}.
\]
Ignoring the electric field for the moment, the particle trajectory satisfies the Lorentz force law. {Choose units such that the mass and magnitude of the charge of the particles is unity and that the particles are positively charged:}
\[
\dot{\mathbf{x}} = \mathbf{v},
\qquad
\dot{\mathbf{v}} = \mathbf{v}\times \mathbf{B}.
\]
Since the Lorentz force is orthogonal to the velocity, the kinetic energy is conserved:
\[
\frac{\mathrm{d}}{\mathrm{d}t}\frac{|\mathbf{v}|^{2}}{2}
=
\mathbf{v}\cdot \dot{\mathbf{v}}
=
\mathbf{v}\cdot (\mathbf{v}\times \mathbf{B})
=
0.
\]

When the magnetic field is spatially uniform, for example $\mathbf{B} = B \hat{z}$, the perpendicular velocity undergoes circular motion with cyclotron frequency $\Omega_c = B$, while the parallel component remains constant. The resulting trajectory is a helix along the magnetic field lines. In this regime, the magnetic moment $\mu$ is exactly conserved.

For slowly varying magnetic fields, the magnetic moment remains approximately conserved. Denoting by $L_B := \frac{B}{|\nabla B|}$, the characteristic magnetic variation length scale and by $\rho = \frac{v_{\perp}}{B}$ the Larmor radius, we define the small parameter
\[
\varepsilon = \frac{\rho}{L_B}.
\]
In the strongly magnetized regime $\varepsilon \ll 1$, standard guiding-center theory yields the adiabatic estimate $\frac{\mathrm{d}\mu}{\mathrm{d}(\Omega_c t)} =\mathcal{O}(\varepsilon)\mu$. Thus, over the fast gyromotion timescale, the magnetic moment varies only weakly and may be treated as an adiabatic invariant.

Averaging over the fast gyromotion therefore yields an effective reduced dynamics along the magnetic field direction. The resulting model evolves only in the longitudinal spatial coordinate and parallel velocity, while the magnetic moment $\mu$ acts as a parameter labeling distinct kinetic slices.

\subsection{Drift-kinetic system}

Under the guiding-center approximation, the multi-species Vlasov equation reduces to the following $1\text{D}1\text{V}$ system parameterized by $\mu$~\cite{JIMENEZ2024magnetic,tyushev2025}. This system is typically termed drift-kinetic model.
\begin{equation}
\label{eq:VP_mirror}
\partial_t f_s
+
v\partial_z f_s
+
\left(
\frac{q_s}{m_s}E(t,z)
-
\mu \partial_z |\mathbf{B}(z)|
\right)\partial_v f_s
=
0,
\end{equation}
where $f_s(t,z,v,\mu)$ denotes the distribution function for species $s\in\{e,i\}$; $z\in\mathbb{R}$ is the longitudinal spatial coordinate; $v\in\mathbb{R}$ is the parallel velocity; $\mu\in\mathbb{R}_{+}$ is the magnetic moment parameter; and $q_s$ and $m_s$ denote the charge and mass of species $s$, respectively.

The effective acceleration contains two distinct contributions: the self-consistent electric field $E(t,z)$; and the magnetic mirror force $-\mu \partial_z |\mathbf{B}(z)|$. Within the reduced formulation, the magnetic moment $\mu$ remains fixed along characteristics and therefore acts as a parameter rather than a dynamical variable.

The electric field is generated self-consistently through a one-dimensional Poisson equation. After normalizing the charges so that
\[
q_i = 1,
\qquad
q_e = -1,
\]
the electrostatic potential $\phi$ satisfies
\begin{equation}
\label{eq:poisson}
\left\{
\begin{aligned}
-\partial_{zz}\phi
+
\partial_z\phi
\frac{\partial_z |\mathbf{B}(z)|}{|\mathbf{B}(z)|}
&=
2\pi |\mathbf{B}(z)|
\iint
\left(
f_i - f_e
\right)
\,\mathrm{d}v\,\mathrm{d}\mu,
\\
\phi(t,-L_z)
=
\phi(t,L_z)
&=
0.
\end{aligned}
\right.
\end{equation}
The electric field is then recovered through
\[
E(t,z) = -\partial_z \phi(t,z),
\]

and the corresponding electric energy as

\[
\mathcal{E}(t) = \frac{1}{2}\int E(t,z)^{2}\,\mathrm{d}z.
\]

We define the following two density quantities that will play an important role throughout the paper: the three-dimensional density, which is given by
\begin{equation}
\label{eq:rho_3d}
\rho_s(t,z)
=
2\pi |\mathbf{B}(z)|
\iint
f_s(t,z,v,\mu)
\,\mathrm{d}v\,\mathrm{d}\mu,
\end{equation}
and the effective longitudinal density, which is given by
\begin{equation}
\label{eq:rho_1d}
\rho_s^{1D}(t,z)
=
\iint
f_s(t,z,v,\mu)
\,\mathrm{d}v\,\mathrm{d}\mu.
\end{equation}

The important distinction between these two definitions arises because increasing the magnetic field strength pushes the field lines closer together. Since particles mostly follow these magnetic field lines, increasing the field strength also increases the particle density, as can be seen from the $|\mathbf{B}(z)|$ term in \eqref{eq:rho_3d}. However, if we integrate out the two spatial dimensions perpendicular to the magnetic field, the corresponding line density in equation \eqref{eq:rho_1d} does not have this dependence on the magnetic field. Since our model only depends on one spatial variable and takes this compression effect into account, it is more natural to consider the one-dimensional line density, which is also the density corresponding to the mass/charge conservation in our drift-kinetic equation.

\subsection{PDE-constrained optimization problem}

The goal of magnetic mirror design is to identify magnetic field configurations that maximize particle confinement over a prescribed time horizon. To this end, we parameterize the magnetic field profile through a neural network representation
\[
|\mathbf{B}(z)| = \mathcal{N}(z;\theta),
\]
where $\theta$ denotes the trainable parameters. The optimization problem is then formulated as
\begin{equation}
\label{eq:objective}
\max_{\theta}\mathcal{J}(\theta)
=
\max_{\theta}
\sum_{s\in\{e,i\}}
\int_{-L_z}^{L_z}
\rho_s^{1D}(T,z;\theta)
\,\mathrm{d}z,
\end{equation}
subject to the reduced drift-kinetic system~\eqref{eq:VP_mirror}--\eqref{eq:poisson}.

Here, $\rho_s^{1D}(T,z;\theta)$ denotes the longitudinal particle density \eqref{eq:rho_1d}, associated with the magnetic field generated by the parameter vector $\theta$. The objective functional therefore measures the total amount of plasma retained within the computational domain at the final simulation time $T$ within the domain of size $2L_z$.

Throughout this work, gradients of the objective functional with respect to the neural network parameters are computed through reverse-mode automatic differentiation applied directly to the discretized numerical solver, yielding an end-to-end differentiable optimization framework.

% ==========================================
% 3. Nonlinear effect and loss-cone
% ==========================================

\section{Nonlinear effects and limitations of loss-cone theory}\label{sec:loss_cone}

Classical magnetic mirror confinement theory is primarily derived from single-particle dynamics under prescribed magnetic fields. Within this framework, confinement behavior is a function of the mirror ratio (the ratio between the largest and smallest magnetic field) and the associated loss-cone criterion, which predicts the fraction of particles expected to remain trapped inside the device.

However, as we demonstrate in Section~\ref{sec:forward}, the self-consistent electrostatic field generated through the Vlasov--Poisson coupling can qualitatively alter the confinement dynamics. In particular, particles with small magnetic moment, which are only weakly influenced by the magnetic mirror force, may nevertheless remain confined due to the emergence of an electrostatic trapping mechanism. Consequently, confinement can no longer be characterized solely through single-particle mirror dynamics.

The objective of this section is to examine the limitations of the classical loss-cone predictions in the nonlinear self-consistent regime. We first investigate the confinement mechanism generated by the self-consistent electric field through forward simulations in Section~\ref{sec:forward}. Then, in Section~\ref{sec:losscone}, we review the classical loss-cone prediction derived from linear transport dynamics and finally compare these theoretical predictions against fully coupled numerical simulations in Section~\ref{sec:loss_cone_pitfall} to demonstrate the breakdown of the loss cone argument.

\subsection{Forward simulations and electrostatic confinement}\label{sec:forward}

The self-consistent electrostatic field generated through the Poisson coupling can fundamentally alter the confinement behavior of the reduced Vlasov system. In particular, we observe the emergence of confinement mechanisms that cannot be explained solely through classical magnetic mirror arguments based on single-particle trajectories and mirror ratios. To illustrate this effect, we first numerically investigate a simplified electron-only regime and subsequently consider the fully coupled multi-species system. The numerical methods used for these simulations are summarized in Section~\ref{sec:computational_setup}.

\begin{figure}[!htb]
    \centering
    \begin{subfigure}[c]{0.29\textwidth}
    \centering
    \includegraphics[width=1.0\linewidth]{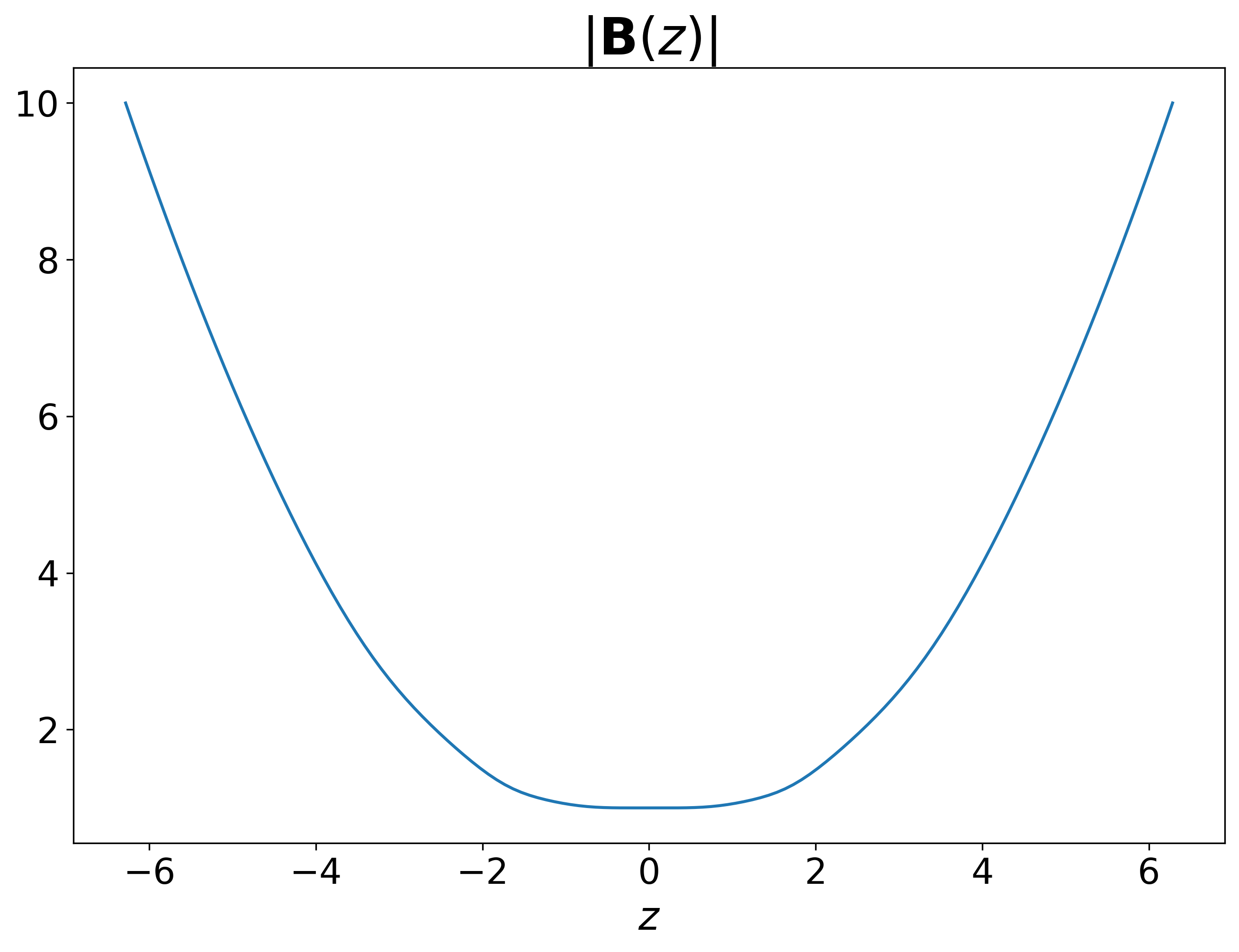}  
    \caption{Single-well magnetic profile}
    \label{fig:intro_init_b}
    \end{subfigure}
    \begin{subfigure}[c]{0.69\textwidth}
    \centering
    \includegraphics[width=1.0\linewidth]{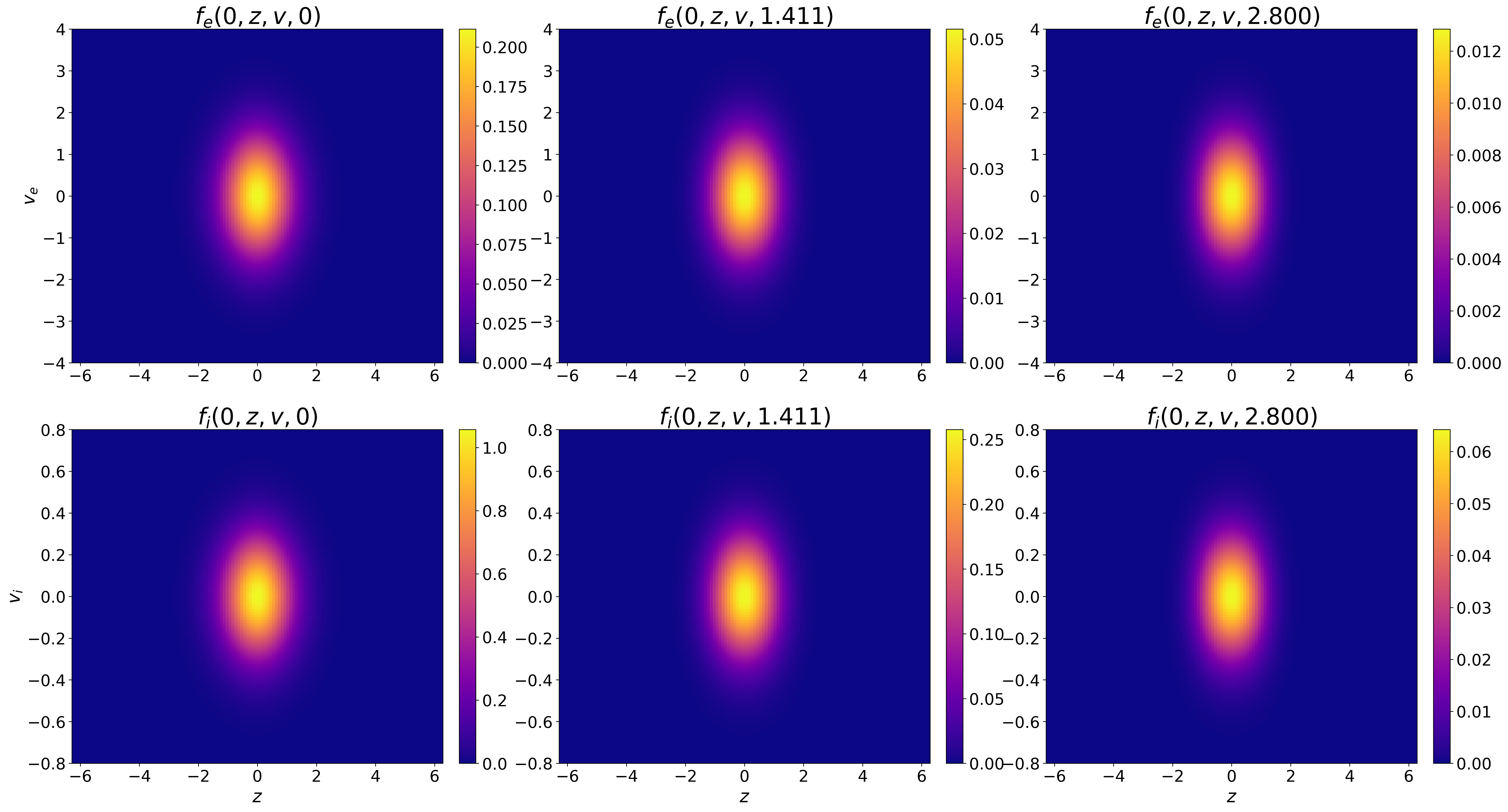}
    \caption{Initial phase-space distributions for electrons (top) and ions (bottom) generated from~\eqref{eq:VP_mirror_init} at three different $\mu$-slices.}
    \label{fig:intro_init_f}
    \end{subfigure}
    \caption{Initial setup for $|\mathbf{B}(z)|$ (left) and $f_{s}(0,z,v,\mu)$ (right).}
    \label{fig:intro_init}
\end{figure}

In both simulations, we initialize the distribution functions using spatially localized Maxwellian-type profiles:
\begin{equation}
\label{eq:VP_mirror_init}
f_s(0,z,v,\mu)=\left[D\exp\left(-\frac{(z-z_c)^2}{2\sigma_z^2}\right)\right]
\left[
\frac{m_s^{3/2}|\mathbf{B}(z)|}{\sqrt{2\pi}}
\exp\left(
-\frac{m_s(v^2+2|\mathbf{B}(z)|\mu)}{2}
\right)
\right],
\end{equation}
where $\sigma_z^2$ controls the spatial variance, $z_c$ denotes the center of the plasma distribution, and $D$ is a normalization constant chosen so that the total initial particle density equals one. Integrating with respect to $v$ and $\mu$ yields
\[
\rho_e^{1D}(0,z)=\rho_i^{1D}(0,z)=D\exp\left(
-\frac{(z-z_c)^2}{2\sigma_z^2}
\right)\,,
\]
so that the initial longitudinal density profile is Gaussian. The initial densities for ions and electrons are chosen identical to ensure charge neutrality at $t=0$. The corresponding initial phase-space distributions are displayed in Figure~\ref{fig:intro_init_f}. The numerical parameters are set according to Table~\ref{tab:simulation_setup} in Section~\ref{sec:computational_setup}.

We first consider a simplified electron-only regime where the ion distribution is treated as an stationary neutralizing background so $f_i$ does not evolve. This approximation is valid on timescales that are short enough such that the heavy ions do not have enough time to respond to the fast electrons. Therefore, this approximation isolates the fast electron dynamics and allows us to study the confinement mechanism generated by the self-consistent electric field. In the simulations we set $m_e=1$ and adopt a prototypical magnetic mirror geometry consisting of a single magnetic well, see Figure~\ref{fig:intro_init_b}.

\begin{figure}[!htb]
    \centering
    \includegraphics[width=1.0\linewidth]{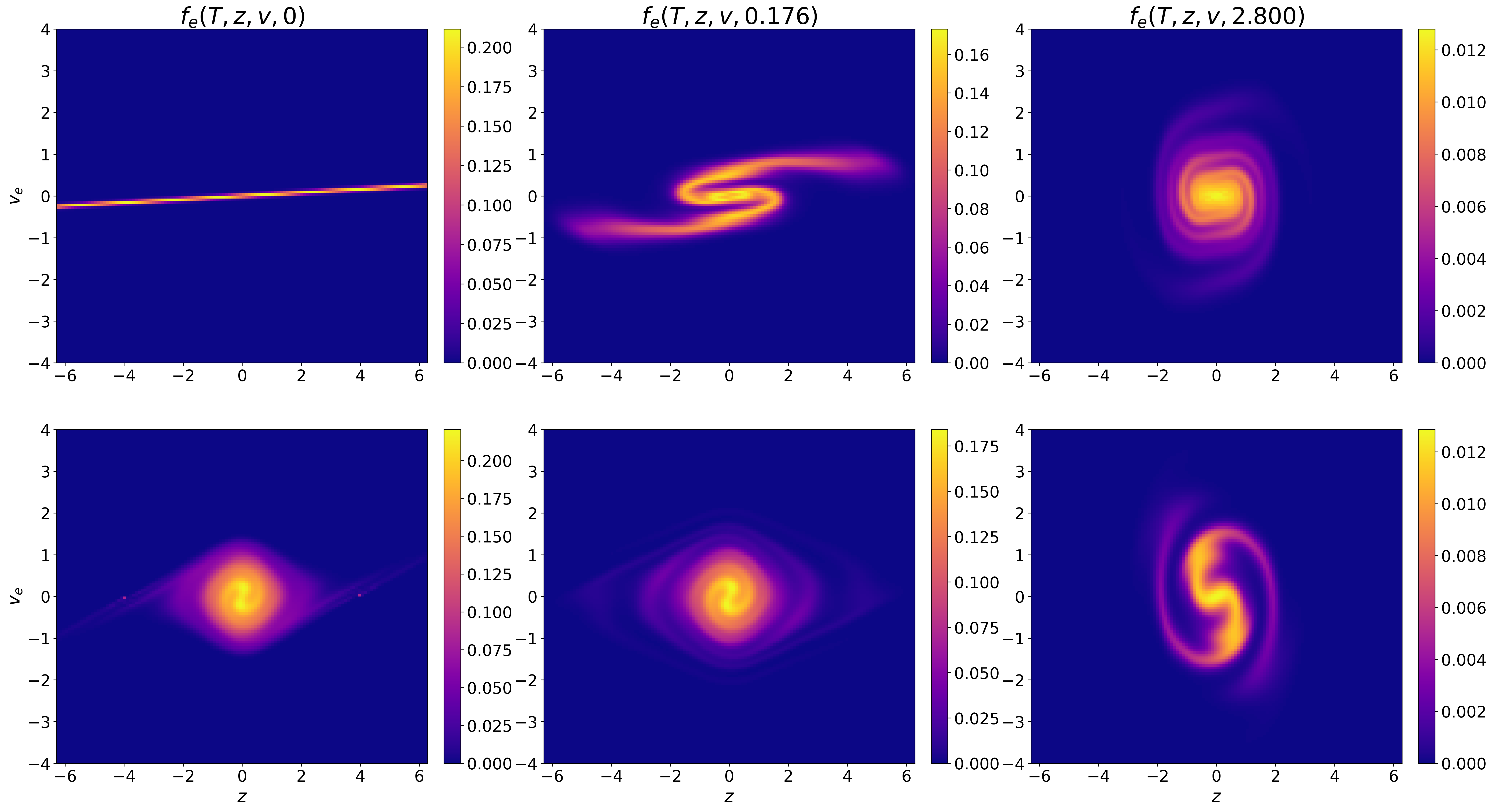}
    \caption{Electron density $f_e(T=25)$ for simulating~\eqref{eq:VP_mirror} with a stationary ion background\ without the presence of $E$ (top) and with the presence of $E$ (bottom). Clearly, the electric field provides another layer of confinement.}
    \label{fig:final_density_E_effect_z-v}
\end{figure}

To examine the role of the self-consistent electric field, we compare simulations with and without the electric field. Figure~\ref{fig:final_density_E_effect_z-v} displays the phase-space distribution at final time $T=25$ for several representative $\mu$-slices. When the self-generated electric field is turned off, the dynamics are effectively linear and particles evolve independently under the magnetic mirror force. In this case, particles with small magnetic moment experience only weak magnetic confinement, since the mirror force $-\mu \partial_z |\mathbf B|$ vanishes as $\mu\to 0$. Consequently, classical single-particle mirror theory predicts that particles with sufficiently small magnetic moment should rapidly escape the confinement region, while particles with larger magnetic moment remain localized. This can be seen the top row of Figure~\ref{fig:final_density_E_effect_z-v}.

In contrast, once the self-consistent electric field is added to the model, the confinement behavior changes significantly. Particles with vanishing magnetic moment ($\mu\approx 0$), which experience negligible magnetic mirror force, nevertheless remain confined. Since magnetic confinement alone cannot explain this phenomenon, the observed trapping mechanism must originate from the nonlinear electrostatic field generated self-consistently through the Poisson equation. The electric field therefore acts as an additional confinement barrier beyond the classical magnetic mirror mechanism, see the bottom row of Figure~\ref{fig:final_density_E_effect_z-v}.

\begin{figure}[!htb]
    \centering
    \includegraphics[width=1.0\linewidth]{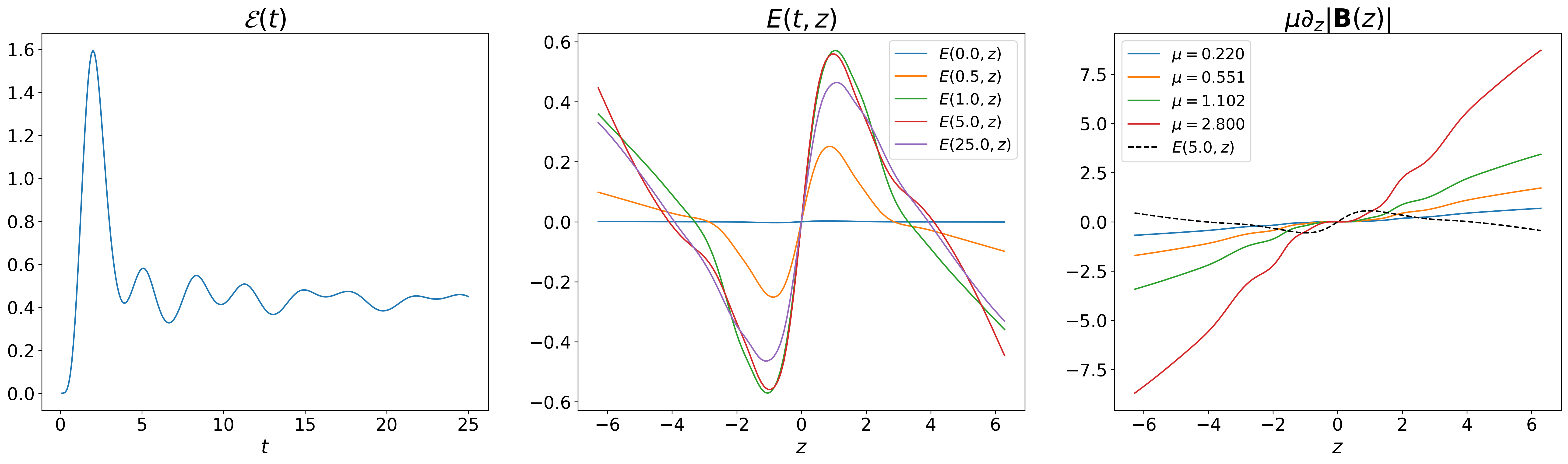}
    \caption{Physical quantities computed from the simulation of the electron drift-kinetic equation with a stationary ion background. From left to right: electric energy $\mathcal{E}(t)$, electric fields $E(t,z)$ for different times $t$ and, mirror force $\mu\partial_{z}|\mathbf{B}(z)|$ for the single-species regime.}
    \label{fig:energies_E_effect_single}
\end{figure}

In Figure~\ref{fig:energies_E_effect_single} we plot the evolution of electric energy in time, different time slices of electric field profile, and the mirror force for multiple choices of $\mu$. While the electric field has different profiles at different time, the rough structure remains the same: with a basin on the left side of the origin and a hill on the right. This structure induces a force that acts in the same direction as the mirror force (i.e.~increases confinement). In the right most panel of the plot, we add the black dotted line as the electric field at time $t=5$, and it is roughly negative of $\mu\partial_z|\mathbf{B}(z)|$ for $\mu\sim 1$. This means, particles with magnetic momentum $\mu\sim 1$, at time $t=5$ sense roughly the same amount of force from magnetic and electric field.

Intuitively, the electric field provides an extra layer of confinement: fast electrons escape the computational (or physical) domain, and since the ions are stationary, a net positive charge builds up in the middle of the domain. This results in an electric field that prevents further leakage. For larger $\mu$, the electric force is negligible compared to the mirror force. However, for particles in the loss cone (small $\mu$), the electric field is the main mechanism that prevents the escape of such particles.

We now turn to the multi-species system. To reduce computational cost while preserving the scale separation between electrons and ions, we adopt the reduced mass ratio
\[m_e = 1,\qquad m_i = 25.
\]
Although this ratio is substantially smaller than the physical proton-electron mass ratio ($m_i/m_e \approx 1836$), it is sufficient to capture the distinct macroscopic timescales of the two species. We point out that reducing the mass ratio in such a manner in order to  reduce computational cost is common in the literature (see, e.g., \cite{vogman2024}).

\begin{figure}[!htb]
    \centering
    \includegraphics[width=1.0\linewidth]{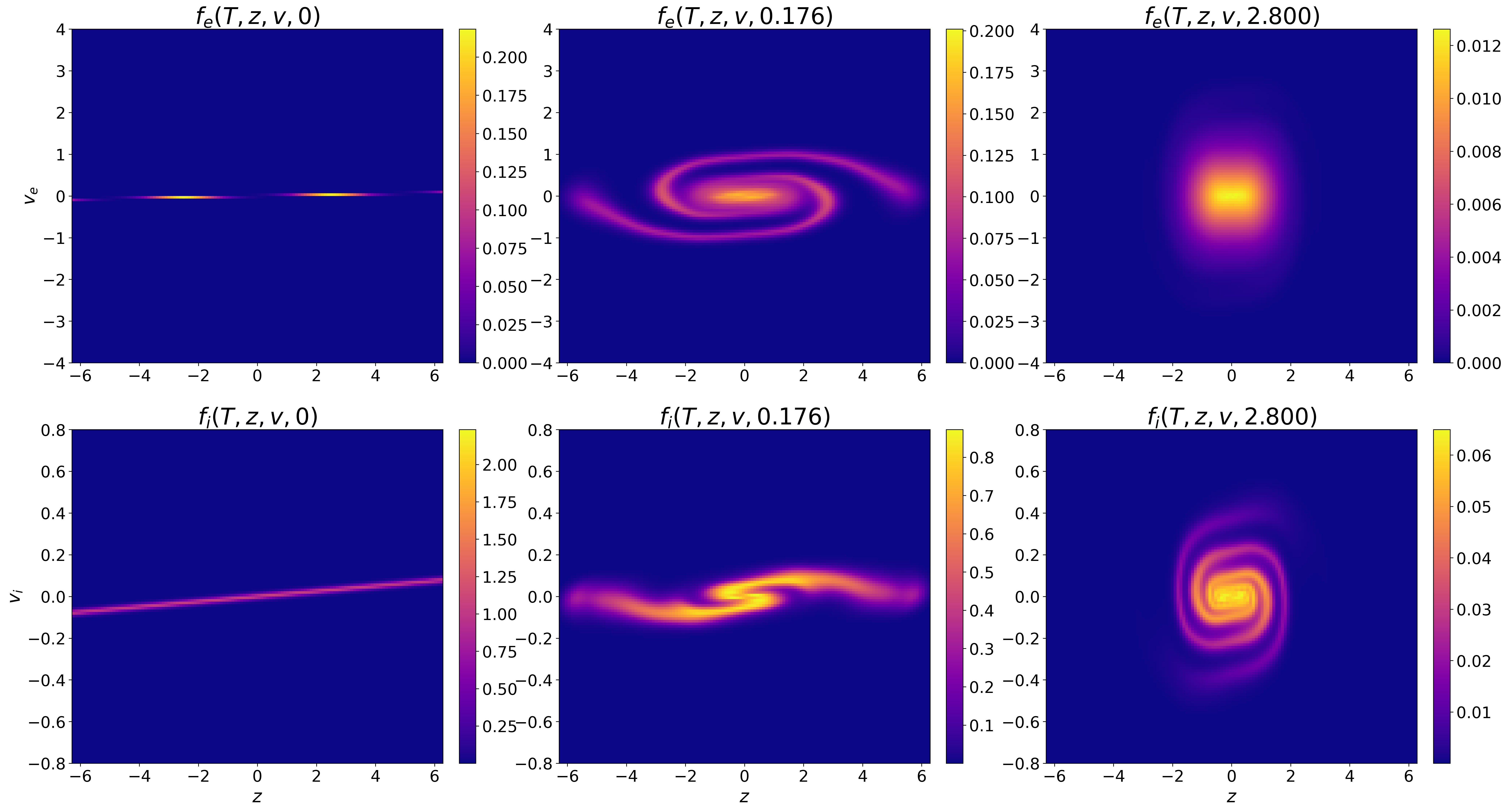}
    \includegraphics[width=1.0\linewidth]{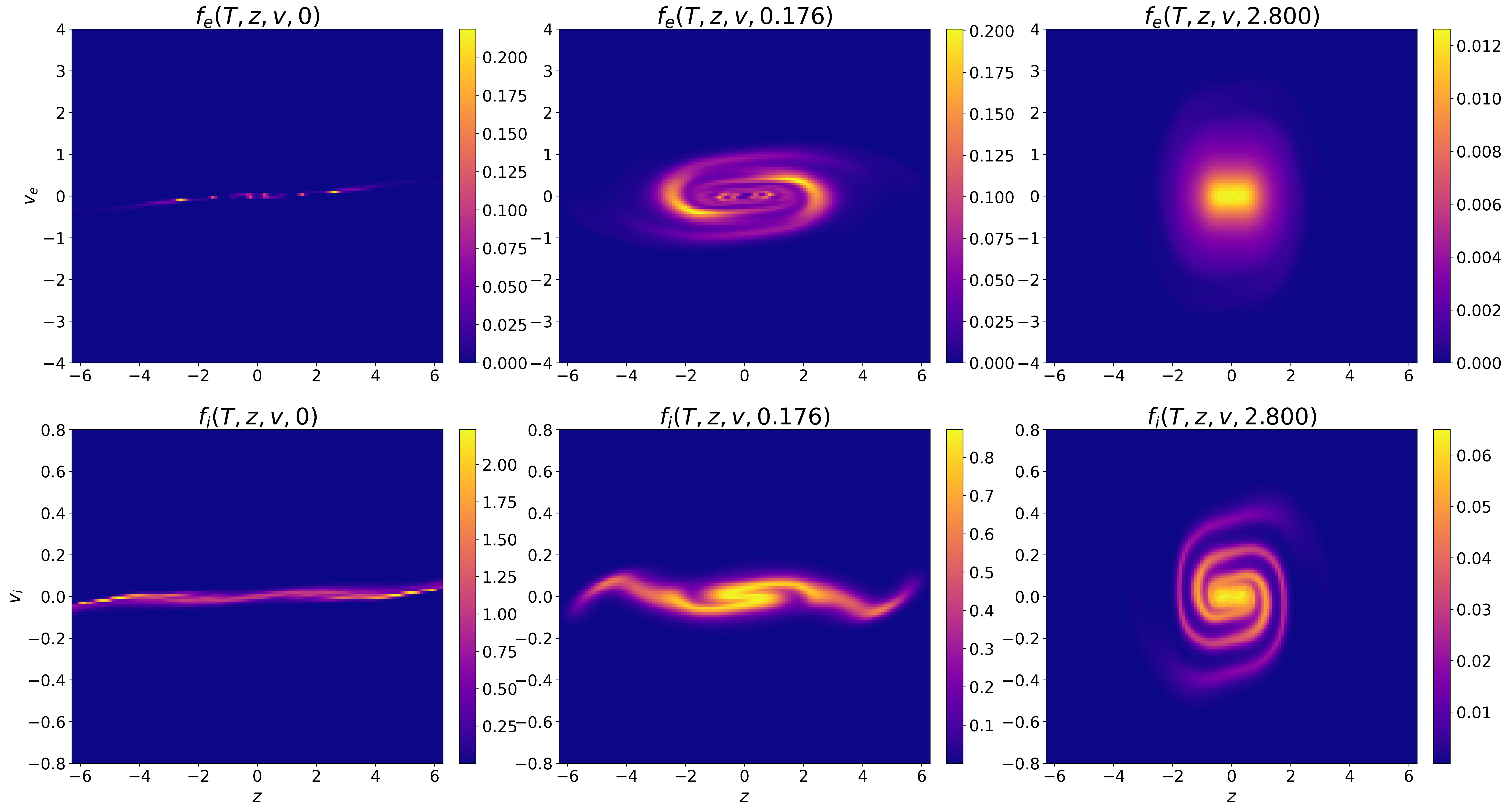}
    \caption{Final particle density $f_s(T=80)$ ($s\in \{e,i\}$) of~\eqref{eq:VP_mirror} without the presence of $E$ (top $2$ rows) and with the presence of $E$ (bottom $2$ rows).}
    \label{fig:intro_final_f_full}
\end{figure}

\begin{figure}[!htb]
    \centering
    \includegraphics[width=1.0\linewidth]{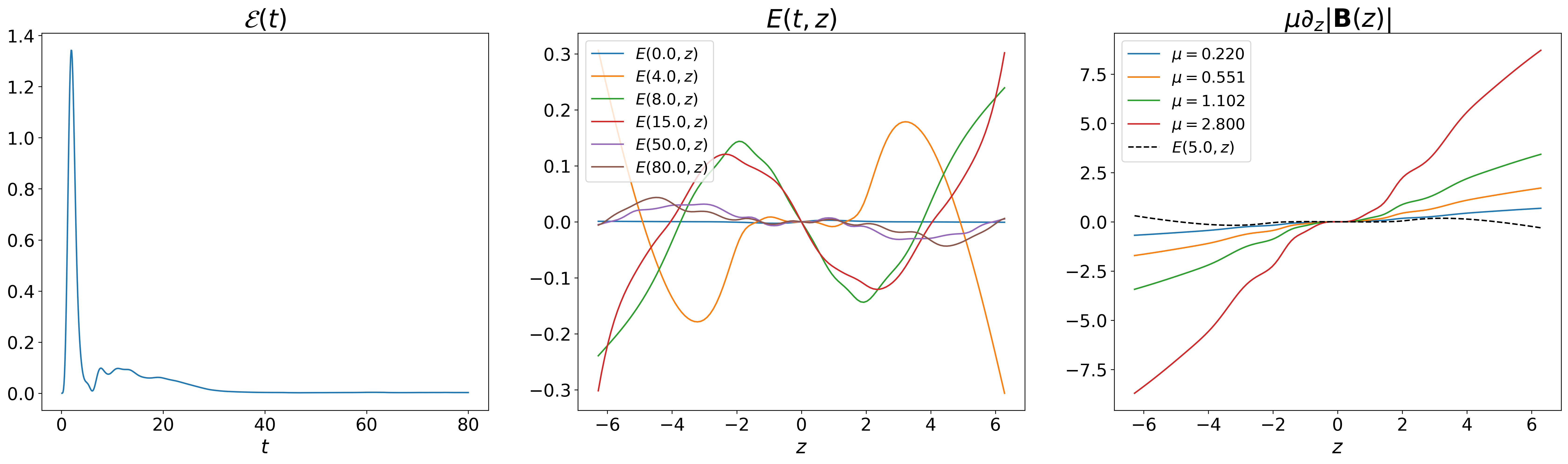}
    \caption{Physical quantities computed from the simulation of the coupled electron-ion drift-kinetic equations. From left to right: electric energy $\mathcal{E}(t)$, electric fields $E(t,z)$ for different times $t$ and, mirror force $\mu\partial_{z}|\mathbf{B}(z)|$ for the full-species regime.}
    \label{fig:energies_E_effect_full}
\end{figure}

Due to the large mass ratio, ions take longer time to escape and saturate, as such  the simulation is performed with a larger time horizon ($T=80$). The resulting phase-space distributions at the final time $T=80$ are shown in Figure~\ref{fig:intro_final_f_full}. The distribution of ions remains roughly unchanged with self-generated electric field turned off or on. The change is more visible for electrons, whose distribution is better confined when the self-generated electric field turned on. Such confinement is more clearly visible for smaller values of $\mu$. Nevertheless, the confinement effect is significantly weaker in comparison to the single-species case (see Figure~\ref{fig:final_density_E_effect_z-v}). Similar to the single-species case, we also plot the electric energy as a function of time, the electric field's evolution in time and the magnetic field for different slices of $\mu$. Compared to Figure~\ref{fig:energies_E_effect_single}, we observe that the strength of electric field in this scenario is weaker, and is only comparable to the magnetic force generated for particles with magnetic momentum $\mu=0.22$, see Figure~\ref{fig:energies_E_effect_full}.

\subsection{Classical loss-cone prediction}\label{sec:losscone}
Classical magnetic mirror theory predicts particle confinement through single-particle dynamics under a prescribed magnetic field. In this framework, particles with sufficiently large parallel velocity are unable to reverse direction before reaching the boundary and therefore escape the confinement region. The corresponding set of escaping trajectories forms the so-called \emph{loss cone} in velocity space.

When the self-consistent electric field is suppressed, namely $E \equiv 0$, the reduced Vlasov equation~\eqref{eq:VP_mirror} becomes a linear transport equation in which particles evolve independently under the magnetic mirror force. Consequently, the confinement properties can be deduced directly from single-particle dynamics by exploiting the conservation of the magnetic moment and kinetic energy.

Let
\[
R_m := \frac{B_{\max}}{B_{\min}}
\]
denote the mirror ratio, where $B_{\max}$ and $B_{\min}$ represent the maximum and minimum magnetic field strengths, respectively. Since the magnetic moment $\mu = \frac{v_\perp^2}{2B}$ is conserved along particle trajectories, we obtain
\begin{equation}
\label{eqn:particle_R_v}\mu = \frac{v_{\perp,\max}^2}{B_{\max}} = \frac{v_{\perp,\min}^2}{B_{\min}}\,,\quad\Rightarrow\quad v_{\perp,\max}^{2}=R_m v_{\perp,\min}^{2}\,.
\end{equation}
Meanwhile, the conservation of kinetic energy gives
$v_{\parallel,\max}^{2} +v_{\perp,\max}^{2}=v_{\parallel,\min}^{2}+v_{\perp,\min}^{2}$. Substituting the magnetic moment relation into~\eqref{eqn:particle_R_v} yields:
\[
v_{\parallel,\max}^{2}=v_{\parallel,\min}^{2}-(R_m-1)v_{\perp,\min}^{2}\,.
\]

A particle is reflected before reaching the boundary if its parallel velocity vanishes at some point along the trajectory, namely $v_{\parallel,\max}^{2}=0$. Introducing the pitch angle $\theta$ through $\sin(\theta) = \frac{v_\perp}{|\mathbf v|}$, the reflection criterion becomes
\[
\sin^2(\theta)
\geq
\frac{1}{R_m}.
\]
Namely, particles escape when their pitch angles:
\[
\theta < \theta_c\,,\quad\text{where}\quad \theta_c
=
\arcsin\left(\frac{1}{\sqrt{R_m}}\right)\,,
\]
where $\theta_c$ is termed the critical pitch angle. The corresponding region in velocity space is referred to as the \emph{loss cone}. Intuitively, particles with sufficiently small pitch angle possess a dominant velocity component parallel to the magnetic field and therefore escape before magnetic reflection can occur.

When the initial velocity distribution is {Maxwellian}, one may explicitly compute the fraction of particles predicted to remain trapped. Integrating over the complement of the loss cone yields
\begin{equation}
\label{eq:F_trapped}
F_{\mathrm{trapped}}
=
\frac{1}{4\pi}
\int_{\theta_c}^{\pi-\theta_c}
\int_0^{2\pi}
\sin(\theta)
\,\mathrm d\phi\,\mathrm d\theta
=
\cos(\theta_c)
=
\sqrt{\frac{R_m-1}{R_m}}.
\end{equation}

Observe that:
\begin{itemize}
    \item if $R_m=1$ (no magnetic mirror), then
    \[
    F_{\mathrm{trapped}}=0,
    \]
    meaning all particles escape;
    
    \item if $R_m\to\infty$ (perfect mirror), then
    \[
    F_{\mathrm{trapped}}\to 1,
    \]
    meaning all particles remain confined.
\end{itemize}

For the magnetic profile used in the simulations of Section~\ref{sec:forward}, the mirror ratio is  set to be $R_m = 10$, thus using the arguments above one predicts
\[
F_{\mathrm{trapped}}
=
\sqrt{\frac{R_m-1}{R_m}}
\approx 0.95.
\]

In addition to predicting the fraction of trapped particles, the classical single-particle picture also provides estimates for the characteristic escape times of different species. Since parallel transport is dominated by the streaming term in~\eqref{eq:VP_mirror}, particles with larger parallel velocity escape more rapidly from the computational domain. Using the same numerical setup as in Section~\ref{sec:forward}, we have $[-L_z,L_z]=[-2\pi,2\pi]$, and the characteristic transit time scales as $t_{\mathrm{esc}} \sim \frac{L_z}{|v|}$. 

\begin{itemize}
    \item  For electrons with maximal parallel velocity $|v_e|=4$, the earliest escape occurs at approximately $t_{\mathrm{esc}}^{(e)} = \frac{2\pi}{4} = \frac{\pi}{2}
\approx 1.57$.
\item For ions, the larger mass reduces the characteristic thermal velocity by a factor $\sqrt{\frac{m_i}{m_e}} = 5$, yielding substantially larger escape times. We set $|v_i|=4/\sqrt{m_i/m_e}$ and the fastest ions do not reach the boundary until approximately $t_{\mathrm{esc}}^{(i)} = \frac{5\pi}{2} \approx
7.85$.
\end{itemize}
Since ions are much heavier and thus slower, they escape at a much larger time horizon, and the mass ratio dictates the pronounced difference of escape dynamics between electrons and ions.

Observe that the classical loss-cone prediction depends solely on the magnetic geometry through the mirror ratio $R_m$. In particular, the derivation neglects modifications of the effective particle dynamics induced by self-consistent electrostatic fields and collective plasma interactions.

\subsection{Breakdown of the loss cone argument}
\label{sec:loss_cone_pitfall}

The loss-cone criterion reviewed in the previous subsection predicts confinement is solely determined by the mirror ratio $R_m$ and therefore attributes trapping entirely to magnetic reflection. However, the simulations of Section~\ref{sec:forward} suggest the existence of an additional confinement mechanism generated by the self-consistent electrostatic field and that its effect depends on many factors, including whether the ions can be regarded as stationary, and the mass ratio between ions and electrons. In general, particles with small magnetic moment, which experience only weak magnetic mirror forces, are more affected by the self-consistent electric field. We now investigate quantitatively how this nonlinear effect modifies the predictions of classical loss-cone theory.

\begin{figure}[!htb]
\centering
\includegraphics[width=0.5\linewidth]{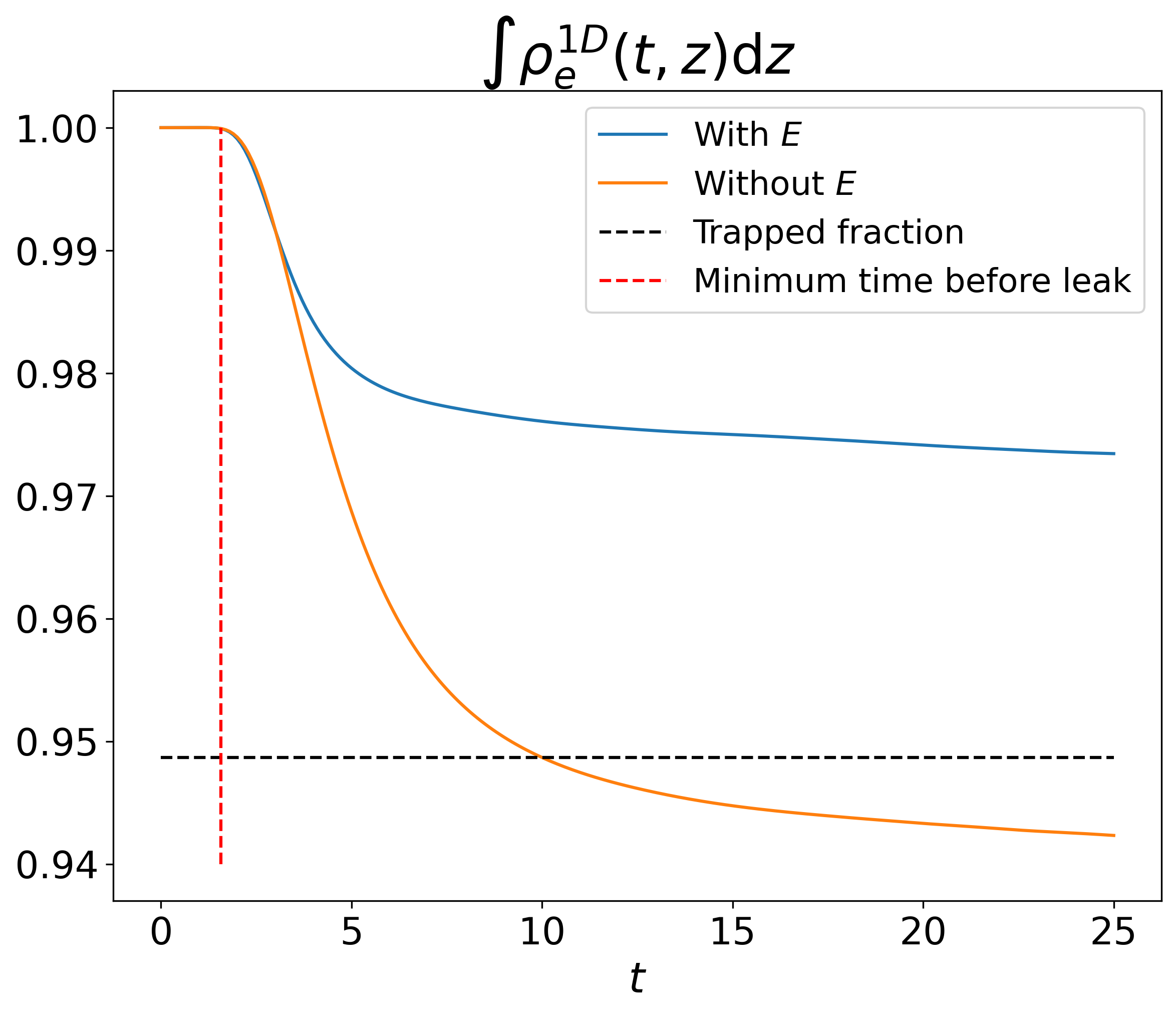}
\caption{Retained electron mass as a function of time for simulations with and without the self-consistent electric field when ions are stationary. The nonlinear electrostatic coupling increases the amount of confined plasma.}
\label{fig:int_rho_E_effect}
\end{figure}

For the electron-only system, Figure~\ref{fig:int_rho_E_effect} reports the retained particle mass
\[
M_e(t)=\int_{-L_z}^{L_z}
\rho_e^{1D}(t,z)\mathrm{d}z.
\]
When the electric field is suppressed, the dynamics reduce to the linear transport regime discussed in Section~\ref{sec:losscone}. In this case, the retained mass converges to approximately $94\%$, in agreement with the classical loss-cone prediction~\eqref{eq:F_trapped} for a mirror ratio $R_m=10$. The earliest particle loss is predicted at approximately $t\approx 1.57$, and this agrees well with the simulation also, see red dash line in Figure \ref{fig:int_rho_E_effect}.

Once the self-consistent electric field is restored, the confinement behavior changes significantly. The retained particle mass now saturates above $97\%$, indicating that particles predicted to escape under purely magnetic confinement remain trapped in the nonlinear regime. This observation is consistent with the phase-space distributions shown in Figure~\ref{fig:final_density_E_effect_z-v}, where particles with $\mu\approx 0$ remain localized despite experiencing negligible magnetic mirror force. The additional confinement therefore originates from the electrostatic potential generated self-consistently through the Poisson equation, and it serves as a second layer of confinement. This is a pure nonlinear and collective mechanism.

\begin{figure}[!htb]
\centering
\includegraphics[width=1.0\linewidth]{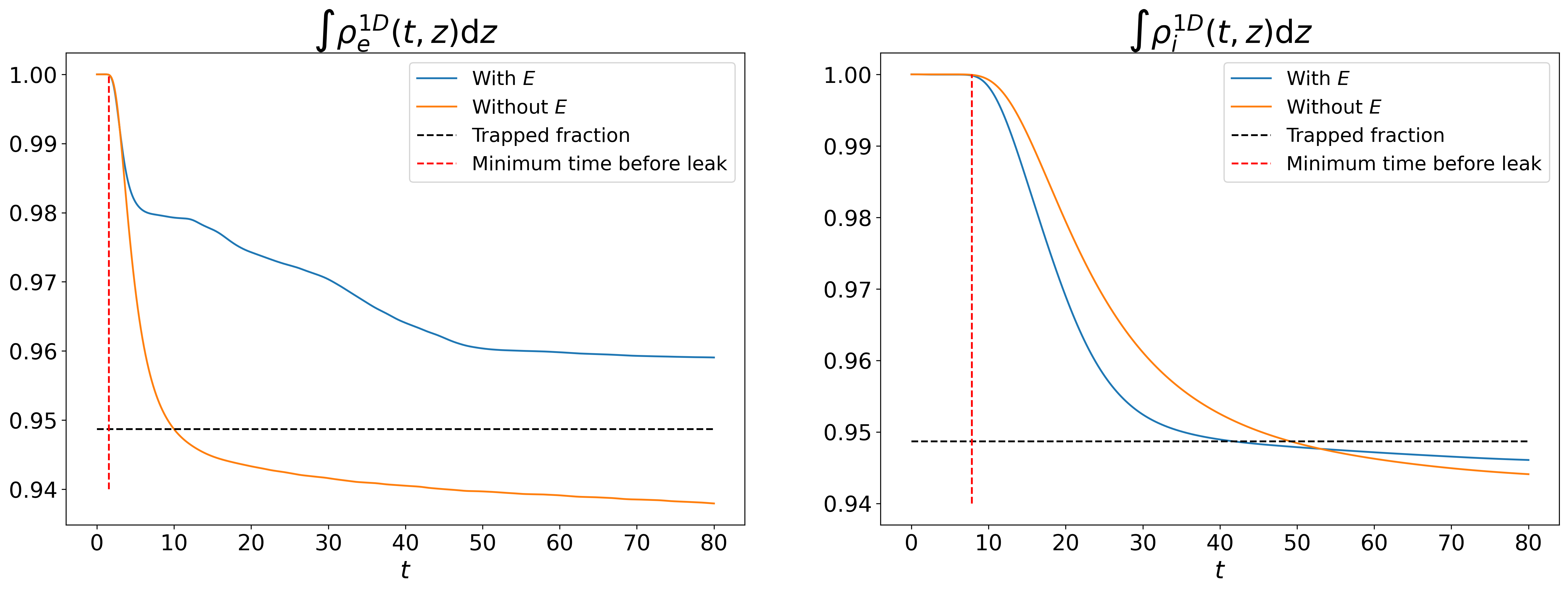}
\caption{Retained particle mass for the fully coupled electron-ion system. Left: electron mass evolution. Right: ion mass evolution. The coupled system exhibits both electrostatic confinement and a pronounced separation between electron and ion escape times.}
\label{fig:int_rho_E_effect_multi}
\end{figure}

We next consider the fully coupled electron-ion system. The corresponding density evolution is reported in Figure~\ref{fig:int_rho_E_effect_multi}. Once again, the initial escape time is expected to be $7.85$ for ions and is plotted as the red dashed line, and this agrees with the simulation well. But we also observe four distinct phases.
\begin{itemize}
    \item \textbf{Initial Electron Escape (roughly $t < 7.85$):} Highly mobile electrons rapidly exit. This phase is when the heavier ions, due to inertia, remain confined and the dynamics of electron is very similar to that of the single-species situation, see comparison with Figure~\ref{fig:int_rho_E_effect}. 
    % During this phase, ions experience faster leakage when electric field is turned on.
    \item \textbf{Ambipolar Confinement (roughly $7.85 < t < 15$):} Electron depletion generates a net positive charge. The resulting electric field accelerates ion expulsion while keeping electrons confined.
    \item \textbf{Field Reversal and Secondary Escape ($15 < t < 50$):} Continuous ion loss alters the self-consistent field, triggering a secondary wave of electron expulsion. At about $t=50$, ion loss is stronger than electron loss, generating a net negative charge, and resulting in a crossover of confining effect with electric field turned on and off.
    \item \textbf{Equilibrium ($t > 50$):} The system relaxes into a quasi-steady state with minimal subsequent mass loss. Net negative charge confines ions, and combined effect of electric and magnetic field confines electrons.
\end{itemize}
This four phase dynamics cannot be drawn from classical loss-cone argument, and the disparity between having self-generated electric field switched on and off is clearly visible comparing the orange and blue lines. Overall, the fraction of both electrons and ions that are confined is significantly larger with the electric field switched on.

Taken together, these results demonstrate that confinement cannot be characterized solely through the mirror ratio. While the classical loss-cone argument captures important aspects of magnetic reflection, the self-consistent electrostatic field introduces additional confinement mechanisms that substantially modify both the amount of retained plasma and the temporal evolution of particle escape. Consequently, confinement performance depends on the full nonlinear dynamics.

% ==========================================
% 4. NUMERICAL Experiments
% ==========================================

\section{Numerical Experiments}
\label{sec:numerics}

So far, we have numerically confirmed that the classical loss-cone theory fails to capture the nonlinear additional confinement mechanisms induced by the electrostatic field. Therefore, the confinement should be also a function of the geometry of external field and not only the mirror ratio $R_m$. 

In this section, we explore how the geometry affects the confinement. Thus, we investigate the PDE-constrained optimization problem introduced in Section~\ref{sec:math_formulation} aiming at finding optimal confinement configurations. This section first summarizes the computational setup and optimization framework, followed by numerical studies of the resulting optimal magnetic mirror configurations. To facilitate verification and further development of this approach, we have made the complete source code, simulation parameters, and optimization configurations publicly available~\footnote{Please see~\url{https://github.com/maguerrap/vp-mirror_model} for the collection of all code and numerical results related to this paper.}.

The simulation parameters used throughout this section are summarized in Table~\ref{tab:simulation_setup}.

\begin{table}[!htb]
\small
    \centering
    \renewcommand{\arraystretch}{1.3} % Adds slight vertical padding for readability
    \begin{tabular}{|c|c|}
        \hline\hline
        \textbf{Parameter / Domain} & \textbf{Values and Setup} \\
        \hline
        Mass and charges & $m_{e} = 1$, $q_{e} = -1$, $m_{i} = 25$, $q_{i} = 1$ \\
        Magnetic mirror & $B_{\min} = 1$, $B_{\max} = 10 \implies R_{m} = 10$ \\
        Temporal domain & $t\in [0,T]$, ($T=25$ single-species, $T=80$ multi-species) \\ % $\Delta t = 0.1$
        Spatial domain & $z\in [-L_{z},L_{z}]$ with $L_{z} = 2\pi$ \\
        Velocity domain & $v_{e} \in [-v_0,v_0]$ ($v_0 = 4$), and $v_{i} \in [-v_0/\sqrt{m_{i}}, v_0/\sqrt{m_{i}}]$ \\
        Magnetic moment & $\mu \in [0,\mu_{0}]$, bounded by tolerance $e^{-\mu_{0}B_{\max}} = 10^{-12}$ \\
        \hline\hline
    \end{tabular}
    \caption{Parameters and domain setup for the numerical simulations.}
    \label{tab:simulation_setup}
\end{table}

\subsection{Computational setup}
\label{sec:computational_setup}

The reduced drift-kinetic system is discretized on a uniform phase-space grid in $(z,v,\mu)$ and advanced in time using Strang operator splitting~\cite{Einkemmer2014}. The resulting transport equations are solved using a conservative semi-Lagrangian method. Since the numerical discretization is not the focus of this work, implementation details are given in  Appendix~\ref{sec:numerical_scheme}.

The principal computational challenge lies in the optimization of the magnetic field geometry. Following the formulation of Section~\ref{sec:math_formulation}, we seek magnetic field profiles that maximize the confinement objective~\eqref{eq:objective}. To this end, the magnetic field is parameterized through a neural network representation
\[
|\mathbf B(z)| = \mathcal N(z;\theta),
\]
where $\theta$ denotes the trainable network parameters. A schematic of the complete neural-network architecture is provided in Appendix~\ref{sec:MLP_architecture}; see Figure~\ref{fig:MLP_architecture}.

Several structural constraints are imposed on the admissible magnetic field profiles. First, the magnetic field is required to remain symmetric with respect to the center of the device. This symmetry is enforced by using the squared distance $(z-z_c)^2$ as the network input. Second, all admissible fields are constrained to share the same mirror ratio
\[
R_m=\frac{B_{\max}}{B_{\min}}=10.
\]
To enforce this constraint, the raw network output is evaluated on the spatial grid, normalized, and scaled to the interval $1 \le |\mathbf B(z)| \le 10$.

This normalization plays an important role in the interpretation of the optimization results. Since all candidate magnetic fields possess the same mirror ratio, any improvement in confinement cannot be attributed to a stronger magnetic mirrors. Instead, performance gains must arise from the detailed spatial configuration of the magnetic field and its interaction with the drift-kinetic model.

The optimization problem is solved using gradient-based methods. Within each iteration, we differentiate directly through the full numerical pipeline using reverse-mode automatic differentiation.

Consequently, the optimization procedure can efficiently exploit gradient information from the underlying kinetic dynamics, enabling end-to-end PDE-constrained optimization of magnetic mirror geometries. The overall optimization procedure is summarized in Algorithm~\ref{alg:optimization_loop}.

\begin{algorithm}[!htb]
\small
\caption{Differentiable Optimization of the Magnetic Mirror Profile}
\label{alg:optimization_loop}
\begin{algorithmic}[1]
\Require Initial distribution $f_s(0,z,v,\mu)$ for $s\in \{e,i\}$, initial network weights $\theta_0$, learning rate $\alpha$, total iterations $N_{\text{iters}}$, simulation end time $T$, time step $\Delta t$, grid $\{z_{i},v_{j},\mu_{k}\}_{i,j,k=0}^{N-1}$.
\Ensure Optimized network weights $\theta^*$
\State $\theta \gets \theta_0$
\For{iteration $= 1, 2, \dots, N_{\text{iters}}$}
    \State Evaluate MLP to obtain spatial magnetic field $|\mathbf{B}(z;\theta)|$
    \State Initialize states $f_s \gets f_s(0,z,v,\mu)$ for $s\in\{e,i\}$ according to~\eqref{eq:VP_mirror_init}
    \State $f_e(T),f_{i}(T) \gets \texttt{PDE\_Solver}\left(\{\{z_{i},v_{j}^{(s)},\mu_{k}\}_{i,j,k=0}^{N-1},f_s\}_{s\in\{e,i\}}, \Delta t, \left\lfloor \frac{T}{\Delta t}\right\rfloor, |\mathbf{B}(z;\theta)|\right)$ \Comment{See Algorithm~\ref{alg:pde_solver}}
    \State Compute final density $\rho_s^{1D}(T,z;\theta) = \iint f_s(T,z,v,\mu) \, \mathrm{d}v \, \mathrm{d}\mu$ for $s\in \{e,i\}$
    \State Evaluate objective $\mathcal{J}(\theta)$
    \State Compute gradients $\nabla_{\theta}\mathcal{J}(\theta)$ via reverse-mode automatic differentiation
    \State $\theta \gets \texttt{Adam}(\theta, \nabla_{\theta}\mathcal{J}(\theta), \alpha)$ \Comment{Update network parameters}
\EndFor
\State \Return $\theta$
\end{algorithmic}
\end{algorithm}

\subsection{Numerical results}
\label{sec:numerical_experiments}
We present our numerical results for both electron-only system and the ion-electron coupled system, summarized in the following subsections respectively. For each case, we present our found optimal magnetic field, and provide physical justification.

\subsubsection{Electron-only regime}

We first investigate the electron-only regime, where ions are treated as a stationary neutralizing background. This setting isolates the interaction between magnetic confinement and the nonlinear electrostatic trapping mechanism identified in Section~\ref{sec:forward}.

\begin{figure}[!htb]
\centering
\includegraphics[width=1.0\linewidth]{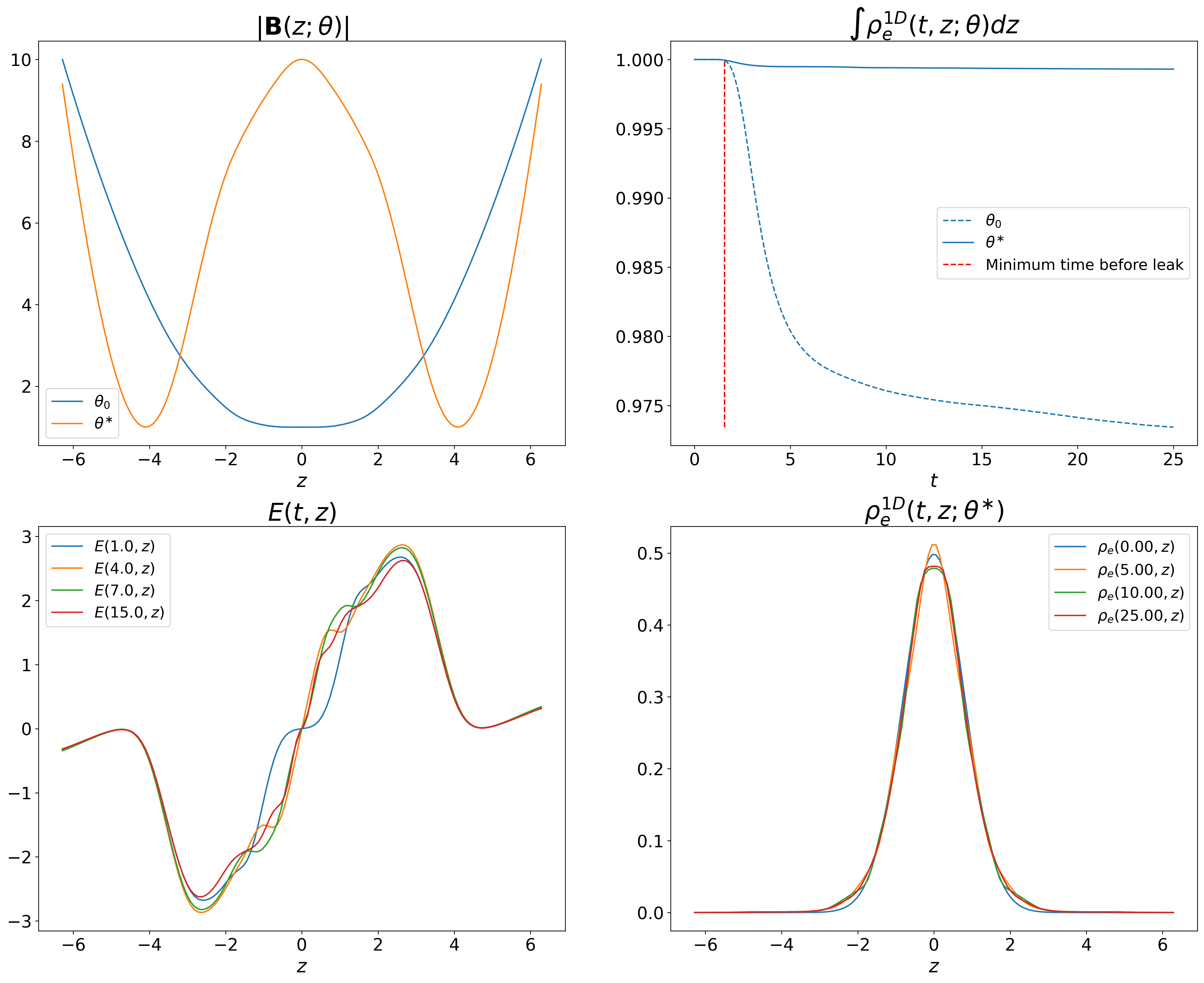}
\caption{Optimization for the electron drift-kinetic equation with stationary ions. Top-left: initial and optimized magnetic field profiles. Top-right: retained particles mass (optimal $\theta^{\ast}$ vs. initial guess $\theta_{0}$) as a function of time. Bottom-left: Electric fields for optimal configuration at four time slices. Bottom-right: Different time-slices for $\rho^{1D}$.}
\label{fig:summary_MLP}
\end{figure}

Figure~\ref{fig:summary_MLP} summarizes the optimization results. Surprisingly, we observe that the optimized field differs qualitatively from the classical magnetic mirror configuration. Rather than forming a single magnetic well with peaks located near the boundaries, the optimizer consistently finds a field profile with a pronounced maximum at the center of the device and two symmetric secondary wells roughly between the middle and the boundaries.

To confirm this is not a consequence of a particular initialization, in Appendix~\ref{sec:robustness_opt} we test and report the robustness of the optimization procedure where we repeat the optimization process using $100$ independent random initializations of the neural network parameters and report the resulting mean magnetic field profile together with a $95\%$ confidence interval. The resulting confidence bands remain narrow throughout the domain and consistently exhibit a pronounced magnetic peak at the center.

Though counter-intuitive, the observed better performance of the double-well field can be explained as follows. There are two layers of this explanation. Firstly, we notice there is a strong magnetic field in the center of the domain. Since the source term in the Poisson equation~\eqref{eq:poisson} scales with $|\mathbf{B}(z)|$, such strong magnetic field also induces a strong electric field in the center of the domain. This is particularly visible if we compare bottom-left of Figure \ref{fig:summary_MLP} with middle of Figure \ref{fig:energies_E_effect_single}, the electric field generated by a single-well field. Physically, since the magnetic field is stronger in the domain center, it pushes the magnetic field lines to be closer, and the cross section of the plasma perpendicular to the magnetic field lines shrinks. This drives up the electric field with the same net charge and accelerates particles with stronger force.

\begin{figure}[!htb]
\centering
\includegraphics[width=1.0\linewidth]{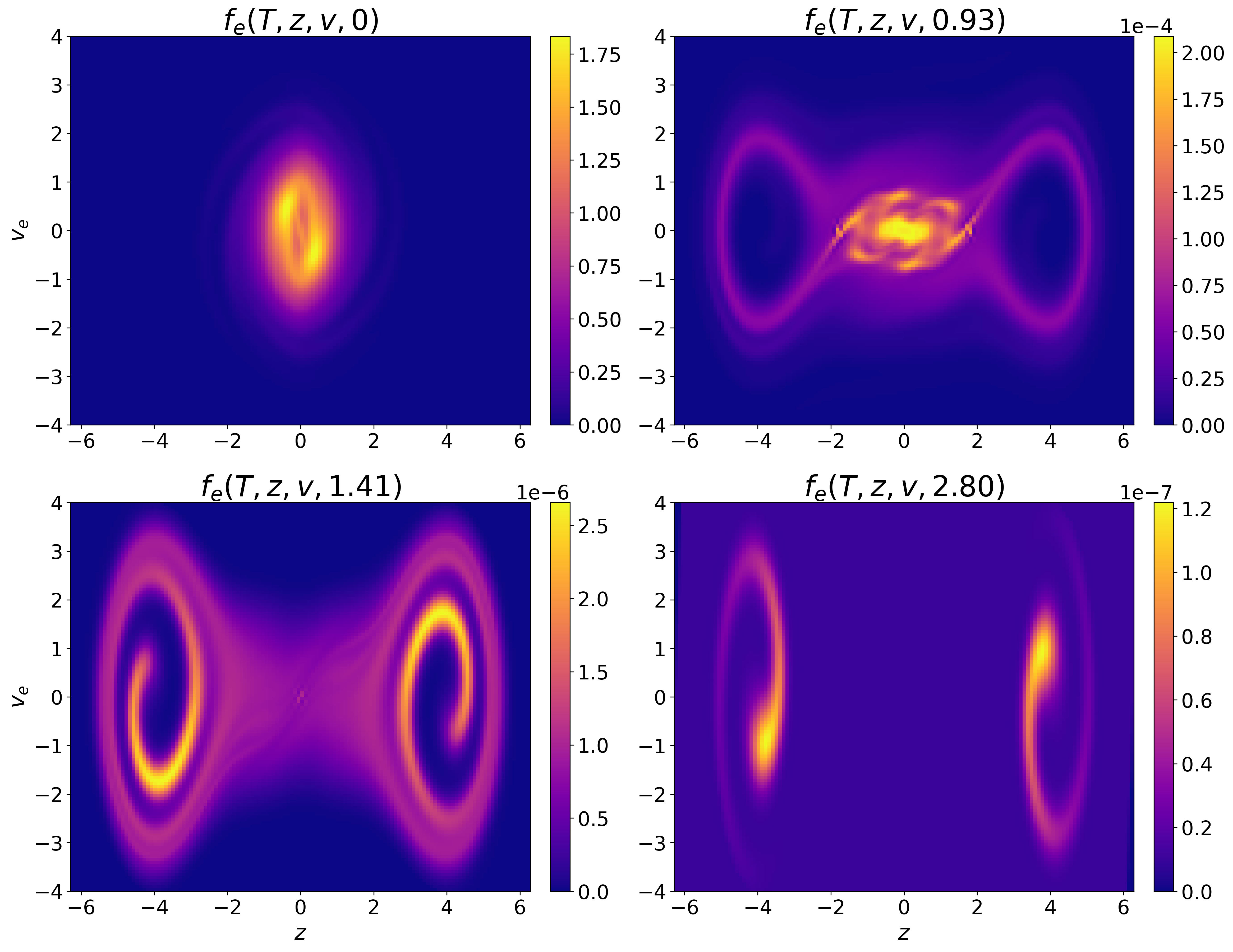}
\caption{Electron phase-space distributions at final time $T=25$ when ions are stationary equipped with the optimized double well magnetic field. The distributions are shown for four $\mu$-slices.}
\label{fig:final_density_MLP}
\end{figure}

Secondly, we also notice the two wells are located symmetrically with respect to the origin. They are also important, and they are in charge of providing confinement to the majority of particles that initially escape from the domain center. These wells create the net positive charge required for the buildup of the electric field. See distribution of particles in Figure \ref{fig:final_density_MLP}. Particles with larger $\mu$ are pushed and confined in the regions of the two wells.

\begin{figure}[!htb]
\centering
 \includegraphics[width=1.0\linewidth]{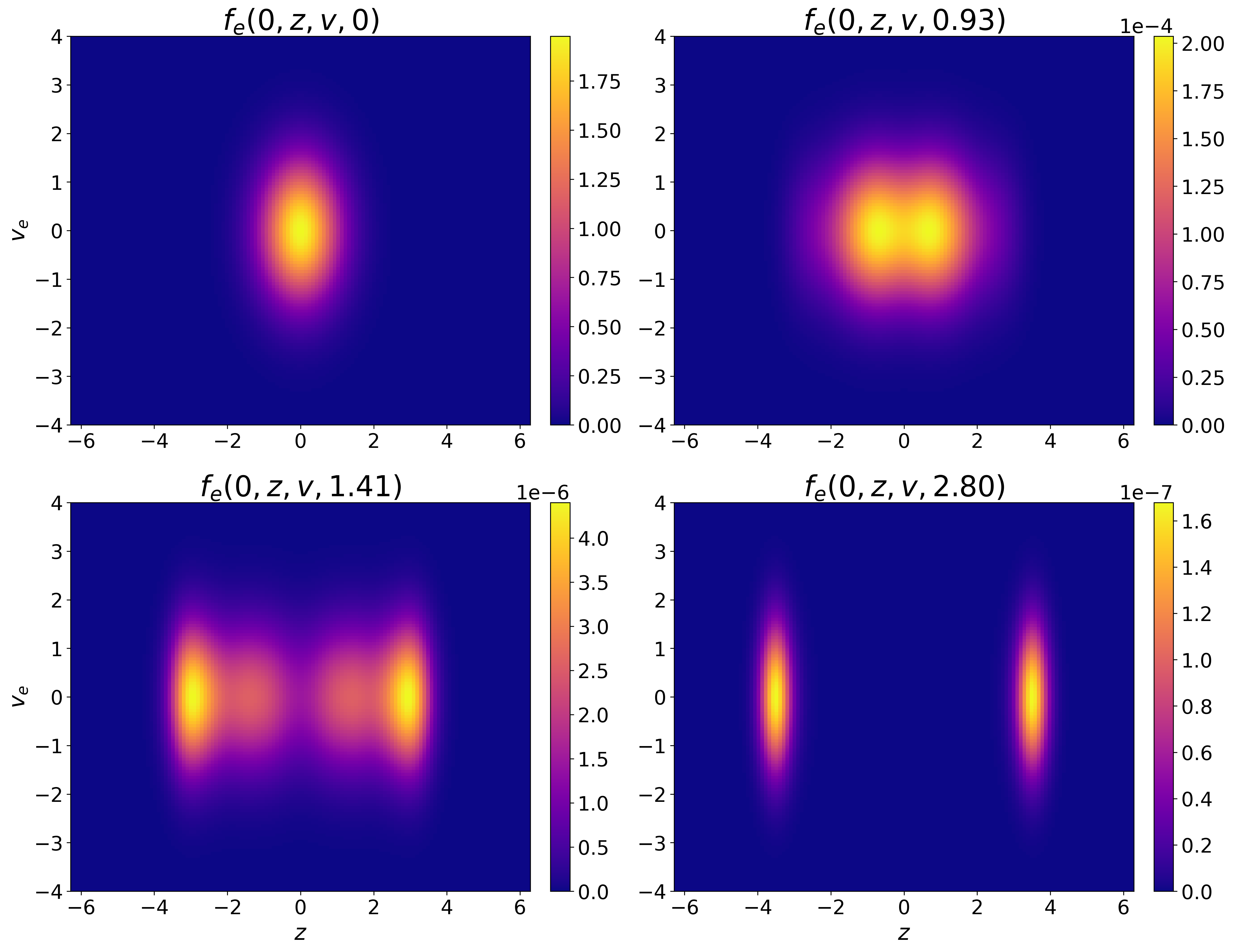}
\caption{Initial phase-space distributions generated from the optimized {double well} magnetic fields {for the drift-kinetic electron simulation with stationary ions.}}
\label{fig:init_density_MLP}
\end{figure}

\subsubsection{Multi-species confinement optimization}

We now consider the fully coupled electron-ion system. Unlike the electron-only regime, the optimization objective must simultaneously account for the confinement of both species and therefore incorporates the complete nonlinear multiscale dynamics.

\begin{figure}[!htb]
\centering
\includegraphics[width=1.0\linewidth]{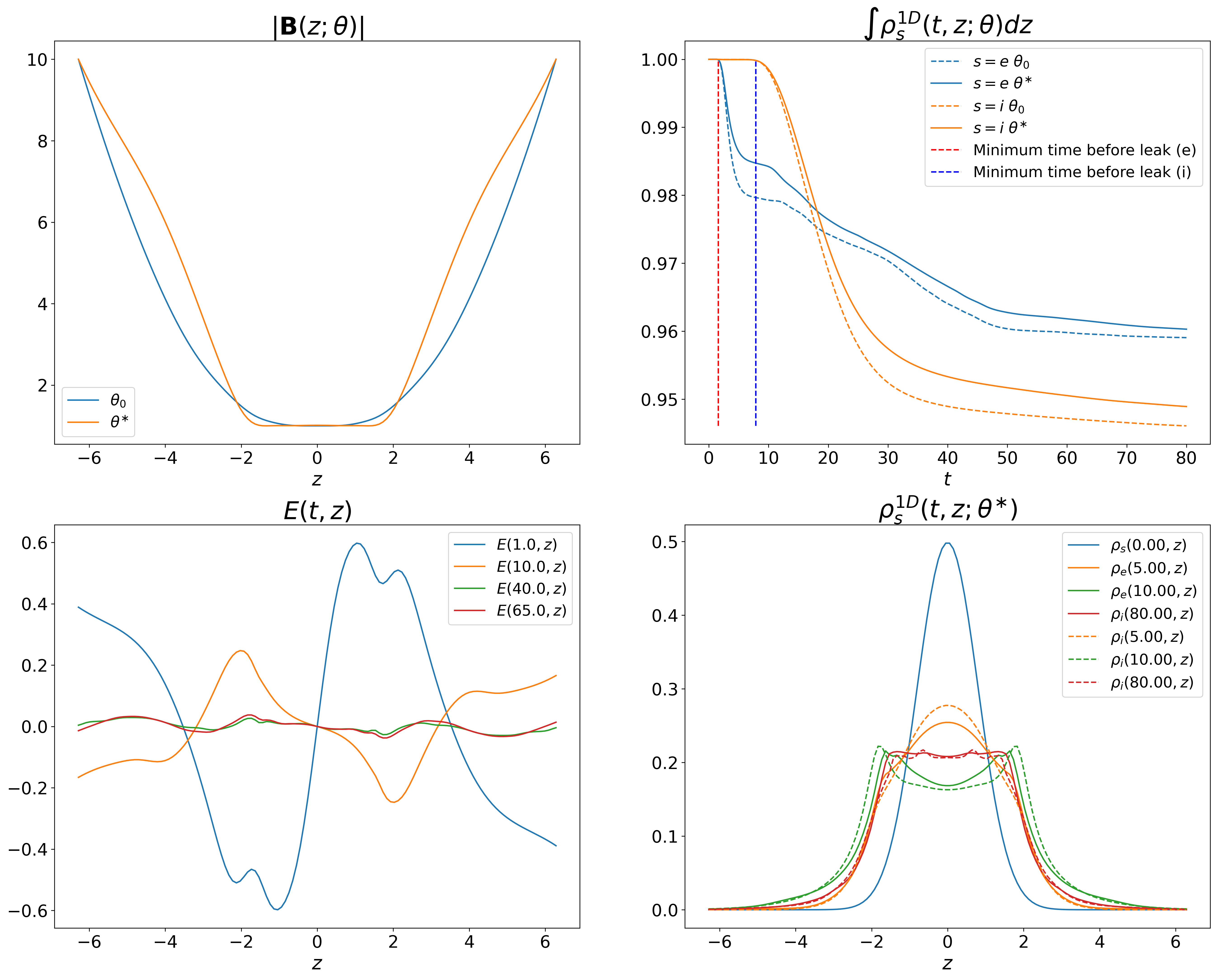}
\caption{Optimization results in the fully coupled electron-ion system. Top-left: initial and optimized magnetic field profiles. Top-right: retained particles mass (optimal $\theta_\star$ vs. initial guess $\theta_0$) as a function of time. Bottom-left: Electric fields for optimal configuration. Bottom-right: Different time-slices for $\rho_{e,i}^{1D}$.}
\label{fig:summary_MLP_full}
\end{figure}

The resulting optimal configurations are shown in Figure~\ref{fig:summary_MLP_full}. The most striking observation is that the preferred magnetic topology changes completely once ion dynamics are introduced. In contrast to the centrally-peaked fields obtained previously, the optimizer now consistently recovers a geometry that closely resembles the classical magnetic mirror configuration: strong magnetic peaks near the boundaries and a broad magnetic well in the center.

As in the single-species case, optimization trajectories originating from different initializations converge to nearly identical magnetic profiles. The robustness is even more pronounced than in the electron-only regime, suggesting the existence of a strongly preferred confinement topology in the electron-ion case.

We perform the same uncertainty quantification study for the electron-ion system. As shown in Appendix~\ref{sec:robustness_opt}, the variance among optimized profiles is even smaller than in the electron-only regime. The optimizer consistently converges toward a boundary-peaked magnetic mirror configuration, indicating an even stronger preference for this topology once ion dynamics are incorporated.

The physical mechanism underlying this different optimization profile transition differs substantially from the electron-only case. Once ion dynamics are included, confinement is no longer dominated by the low-$\mu$ electron population alone. Instead, long-time retention of both species becomes the determining factor. Under these conditions, the optimizer naturally favors a broad central confinement region bounded by strong magnetic mirrors, thereby recovering a configuration reminiscent of traditional mirror devices.

\begin{figure}[!htb]
    \centering
    \begin{subfigure}[t]{1.0\textwidth}
    \centering
    \includegraphics[width=1.0\linewidth]{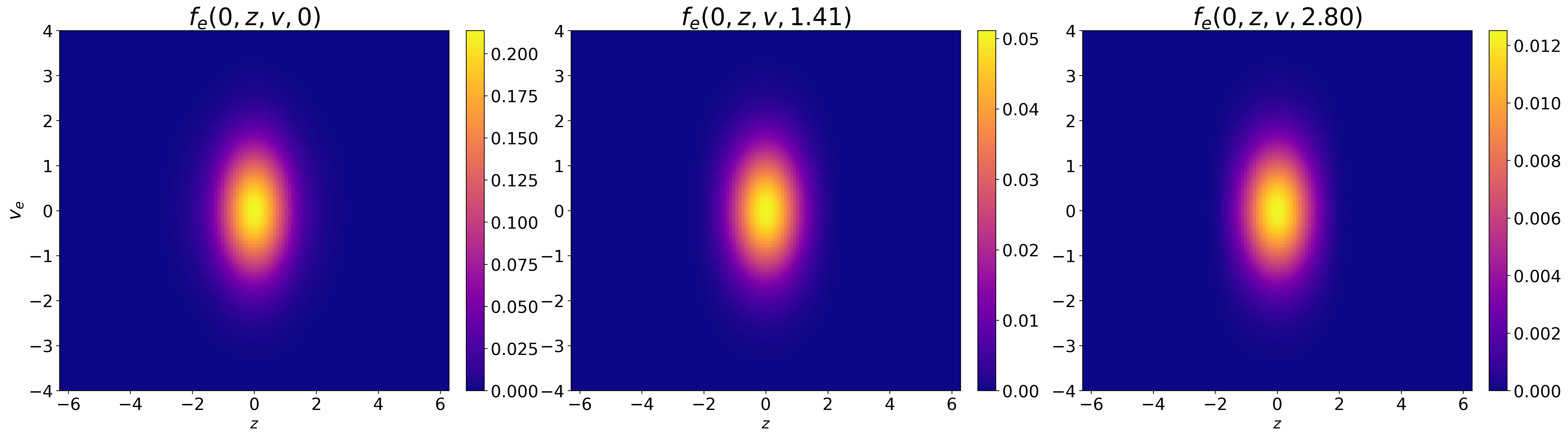}
    \caption{$f_{s} = f_{e}$}
    \label{fig:init_density_MLP_multi_e}
    \end{subfigure}
    \begin{subfigure}[t]{1.0\textwidth}
    \centering
    \includegraphics[width=1.0\linewidth]{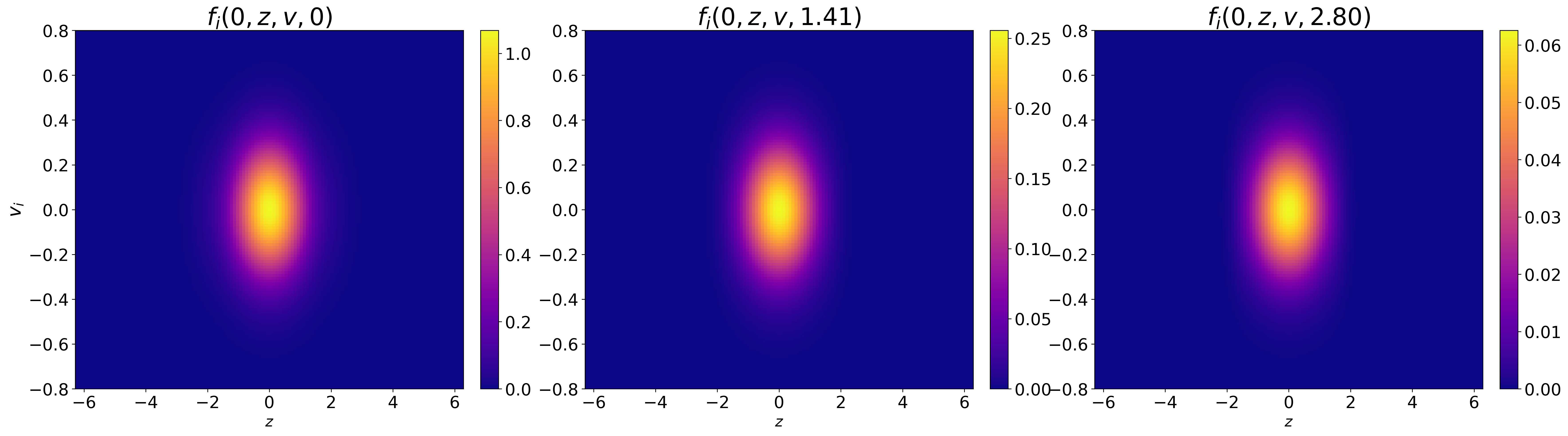}
    \caption{$f_{s} = f_{i}$}
    \label{fig:init_density_MLP_multi_i}
    \end{subfigure}
    \caption{Initial phase-space distributions for the optimized multi-species configuration.}
    \label{fig:init_density_MLP_multi}
\end{figure}

\begin{figure}[!htb]
    \centering
    \begin{subfigure}[t]{1.0\textwidth}
    \centering
    \includegraphics[width=1.0\linewidth]{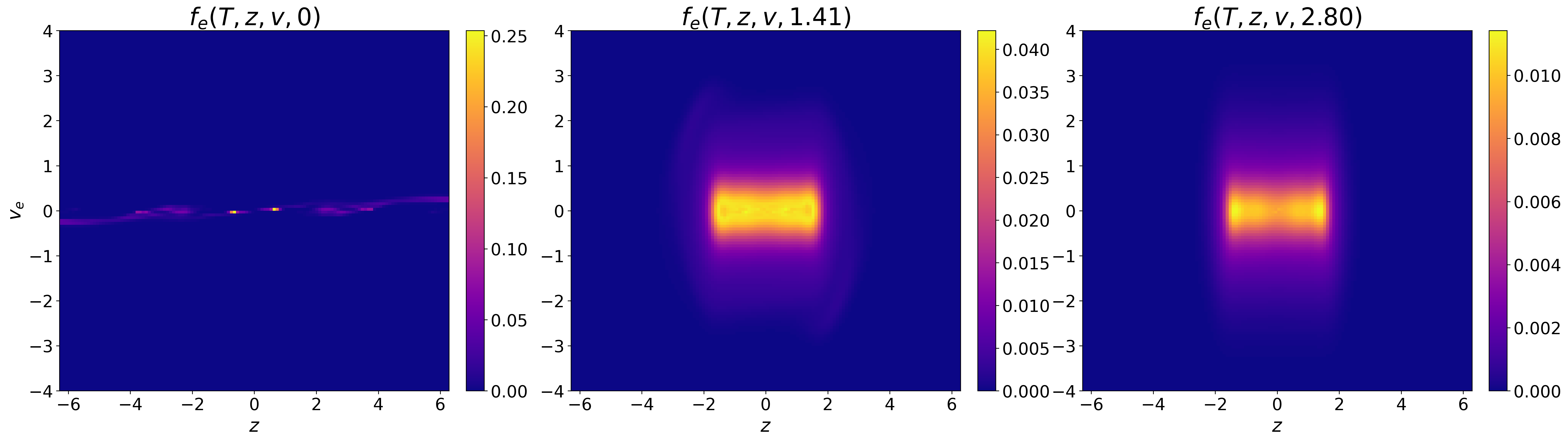}
    \caption{$f_{s} = f_{e}$}
    \label{fig:final_density_MLP_multi_e}
    \end{subfigure}
    \begin{subfigure}[t]{1.0\textwidth}
    \centering
    \includegraphics[width=1.0\linewidth]{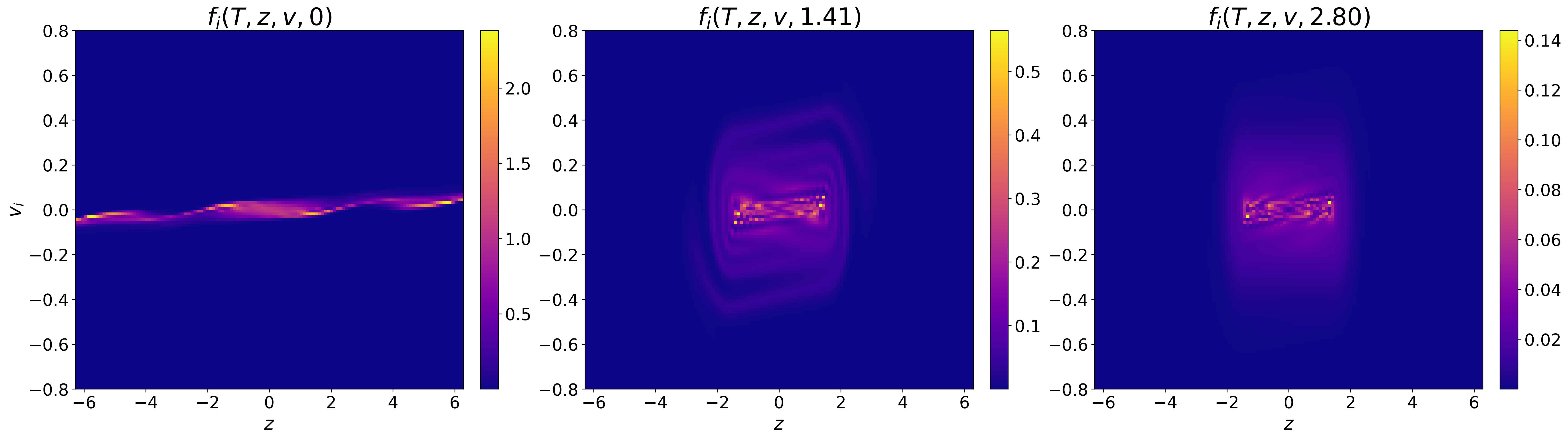}
    \caption{$f_{s} = f_{i}$}
    \label{fig:final_density_MLP_multi_i}
    \end{subfigure}
    \caption{Phase-space distributions at $T=80$ for the optimized multi-species configuration.}
    \label{fig:final_density_MLP_multi}
\end{figure}

Figures~\ref{fig:init_density_MLP_multi} and~\ref{fig:final_density_MLP_multi} further illustrate this behavior. The optimized field concentrates both species near the center of the device while maintaining strong reflection barriers at the boundaries. As a consequence, only particles with magnetic moments extremely close to zero escape during the simulation horizon.

Taken together, these results reveal two distinct confinement strategies discovered through optimization.
\begin{itemize}
    \item In the electron-only regime, the optimizer exploits nonlinear electrostatic trapping and therefore favors centrally-peaked magnetic fields that preserve the dominant low-$\mu$ population.
\item In the fully coupled system, long-time confinement of both species becomes paramount, leading the optimizer to recover a configuration closely resembling the classical magnetic mirror. 
\end{itemize}
The transition between these two optimal topologies highlights the fundamentally different confinement mechanisms operating in the two regimes.

\section*{Acknowledgment}
MG and QL acknowledge the support by ONR Award AWD-100997 and NSF grant DMS-2308440. QL further acknowledges Vilas Associate Award. The work was initiated when MG, QL and LE were visiting Simons Laufer Mathematical Sciences Institute in Berkeley, California, during the Fall 2025 semester to participate in the semester program on ``Kinetic Theory: Novel Statistical, Stochastic and Analytical Methods". This research was funded in part by the Austrian Science Fund (FWF) \url{https://doi.org/10.55776/PAT2937525}.

\printbibliography[heading=bibintoc]

\appendix

\section{Numerical Scheme}\label{sec:numerical_scheme}
In this section, we explain in detail the numerical scheme employed to solve~\eqref{eq:VP_mirror}. We recall that our phase-space is $(z,v^{(s)},\mu)$ for species $s \in \{e,i\}$ which gets discretized into an isotropic grid $\{z_{i},v_{j}^{(s)},\mu_{k}\}_{i,j,k=0}^{N}$ with uniform widths $\Delta z$, $\Delta v^{(s)}$ and $\Delta \mu$. Also, the cell boundaries are denoted by $z_{i\pm 1/2}$ and the time domain is discretized as $\{t^{n}\}_{n=0}^{N_t}$.

\subsection{Finite Volume Scheme}

Let $\mathsf{f}_s^{n}(z_{i},v_{j}^{(s)},\mu_{k}) \approx f_s(t^{n}, z, v^{(s)}, \mu)$ be our numerical approximation now represents the cell-averaged distribution function. A single full time step proceeds in four distinct stages.

First, we perform a half-step spatial advection for both species. We define the discrete spatial primitive function (cumulative mass) evaluated at the cell edges as:
\[
    \mathsf{F}_{s,z}(z_{i+1/2}, v_j^{(s)}, \mu_k) = \sum_{m=0}^{i} \mathsf{f}_{s}^{n}(z_{m}, v_j^{(s)}, \mu_k) \Delta z\,,
\]
where $\mathsf{F}_{s,z}(z_{-1/2}) = 0$. We trace the cell boundaries backward along the characteristics:
\[
    z_{i\pm 1/2}^{\ast} = z_{i\pm 1/2} - v_{j}^{(s)}\Delta t/2\,.
\]
The intermediate distribution is then updated by interpolating the primitive function at the advected edges and computing the flux difference:
\[
    \mathsf{f}_s^{\ast}(z_{i}, v_{j}^{(s)}, \mu_{k}) = \frac{1}{\Delta z}\left[ \mathcal{P}(\mathsf{F}_{s,z})(z_{i+1/2}^{\ast}) - \mathcal{P}(\mathsf{F}_{s,z})(z_{i-1/2}^{\ast}) \right]\,,
\]
where $\mathcal{P}$ represents a shape-preserving Piecewise Cubic Hermite Interpolating Polynomial (PCHIP) interpolation to maintain strict monotonicity. To naturally enforce zero-inflow and open outflow boundaries, $\mathcal{P}(\mathsf{F}_{s,z})$ is strictly clamped to the interval $[0, \mathsf{F}_{s,z}(z_{N-1/2})]$.

Next, we compute the macroscopic 1D densities $\rho_e^{1D}$ and $\rho_i^{1D}$ using the intermediate distributions $\mathsf{f}_e^{\ast}$ and $\mathsf{f}_i^{\ast}$. We use these to solve the Poisson system~\eqref{eq:poisson} for the electrostatic potential $\phi$. We employ a 4th-order finite difference scheme with homogeneous Dirichlet boundary conditions (which we detail in Section~\ref{sec:finite_diff} below). Using these approximations, we evaluate the electric field at the current time step: $\mathsf{E}_{i}^{n+\frac{1}{2}} \approx -\partial_{z}\phi(t^{n+\frac{1}{2}},z_{i})$. We then perform a full-step velocity advection for each species using the same conservative framework. The local characteristic acceleration is computed as $\mathsf{a}_{i,k,s}^{n+\frac{1}{2}} = \frac{q_s}{m_s}\mathsf{E}_{i}^{n+\frac{1}{2}} - \mu_{k}\partial_{z}|\mathbf{B}(z_{i})|$. We trace the velocity boundaries backward:
\[
    v_{j\pm 1/2}^{(s)\ast} = v_{j\pm 1/2}^{(s)} - \mathsf{a}_{i,k,s}^{n+\frac{1}{2}}\Delta t\,.
\]
Defining the velocity primitive function $\mathsf{F}_{s,v}$ generated from $\mathsf{f}_s^{\ast}$ as,
\[
    \mathsf{F}_{s,v}(z_i, v_{j+\frac{1}{2}}^{(s)}, \mu_k) = \sum_{m=0}^{j} \mathsf{f}_{s}^{\ast}(z_i, v_m^{(s)},\mu_{k})\Delta v^{(s)}
\]
then, the full-step velocity advection is computed as:
\[
    \mathsf{f}_s^{\ast\ast}(z_{i},v_{j}^{(s)},\mu_{k}) = \frac{1}{\Delta v^{(s)}}\left[ \mathcal{P}(\mathsf{F}_{s,v})(v_{j+1/2}^{(s)\ast}) - \mathcal{P}(\mathsf{F}_{s,v})(v_{j-1/2}^{(s)\ast}) \right]\,.
\]

Finally, we compute the remaining half-step spatial advection on $\mathsf{f}_s^{\ast\ast}$ exactly as before to complete the time step,

\[
\mathsf{F}_{s,z}(z_{i+\frac{1}{2}},v_{j}^{(s)},\mu_{k}) = \sum_{m=0}^{i}\mathsf{f}_{s}^{\ast\ast}(z_{m},v_{j}^{(s)},\mu_{k})\Delta z\,,
\]
and obtain $\mathsf{f}_s^{n+1}$,
\[
    \mathsf{f}_s^{n+1}(z_{i}, v_{j}^{(s)}, \mu_{k}) = \frac{1}{\Delta z}\left[ \mathcal{P}(\mathsf{F}_{s,z})(z_{i+1/2}^{\ast}) - \mathcal{P}(\mathsf{F}_{s,z})(z_{i-1/2}^{\ast}) \right]\,.
\]

We summarize the generalized numerical integration scheme in Algorithm~\ref{alg:pde_solver}.

\begin{algorithm}[!htb]
\caption{Multi-Species Conservative Semi-Lagrangian \texttt{PDE\_Solver}}
\label{alg:pde_solver}
\begin{algorithmic}[1]
\State \textbf{Input:} Centers and edges of grid $\{z_{i},v_{j}^{(s)},\mu_{k}\}_{i,j,k=0}^{N-1}$ for $s \in \{e,i\}$, initial distributions $\mathsf{f}_s^0$, time step $\Delta t$, final steps $N_t$, magnetic field $|\mathbf{B}(z)|$.
\State \textbf{Output:} Final distributions $\mathsf{f}_s^{N_{t}}(z, v, \mu)$
\State $t \gets 0$
\For{$n = 0,...,N_{t}-1$}
    \For{$s \in \{e,i\}$}
        \State Compute cumulative mass $\mathsf{F}_{s,z}$ from $\mathsf{f}_s^{n}$ \Comment{Half-step $z$ advection}
        \State $z_{i\pm 1/2}^{\ast} \gets z_{i\pm 1/2} - v_{j}^{(s)}\Delta t/2$
        \State $\mathsf{f}_s^{\ast}(z_{i},v_{j},\mu_{k}) \gets \frac{1}{\Delta z} [ \mathcal{P}(\mathsf{F}_{s,z})(z_{i+1/2}^{\ast}) - \mathcal{P}(\mathsf{F}_{s,z})(z_{i-1/2}^{\ast}) ] \quad \forall i,j,k$ 
    \EndFor
    \State Compute net charge density from $\mathsf{f}_e^{\ast}$ and $\mathsf{f}_i^{\ast}$
    \State Solve~\eqref{eq:poisson} via finite differences to obtain $\mathsf{E}_{i}^{n+\frac{1}{2}} \approx E(t+\Delta t/2, z_{i}) \quad \forall i$ \Comment{Self-consistent field}
    \For{$s \in \{e,i\}$}
        \State Compute cumulative mass $\mathsf{F}_{s,v}$ from $\mathsf{f}_s^{\ast}$ \Comment{Full-step $v$ advection}
        \State $\mathsf{a}_{i,k,s}^{n+\frac{1}{2}} \gets \frac{q_s}{m_s}\mathsf{E}_{i}^{n+\frac{1}{2}} - \mu_{k}\partial_{z}|\mathbf{B}(z_{i})|$
        \State $v_{j\pm 1/2}^{(s)\ast} \gets v_{j\pm 1/2}^{(s)} - \mathsf{a}_{i,k,s}^{n+\frac{1}{2}}\Delta t$
        \State $\mathsf{f}_s^{\ast\ast}(z_{i},v_{j}^{(s)},\mu_{k}) \gets \frac{1}{\Delta v} [ \mathcal{P}(\mathsf{F}_{s,v})(v_{j+1/2}^{(s)\ast}) - \mathcal{P}(\mathsf{F}_{s,v})(v_{j-1/2}^{(s)\ast}) ] \quad \forall i,j,k$ 
        \State Compute cumulative mass $F_{s,z}$ from $\mathsf{f}_s^{\ast\ast}$ \Comment{Half-step $z$ advection}
        \State $z_{i\pm 1/2}^{\ast} \gets z_{i\pm 1/2} - v_{j}^{(s)}\Delta t/2$
        \State $\mathsf{f}_s^{n+1}(z_{i},v_{j}^{(s)},\mu_{k}) \gets \frac{1}{\Delta z} [ \mathcal{P}(\mathsf{F}_{s,z})(z_{i+1/2}^{\ast}) - \mathcal{P}(\mathsf{F}_{s,z})(z_{i-1/2}^{\ast}) ] \quad \forall i,j,k$ 
    \EndFor
    \State $t \gets t + \Delta t$
\EndFor
\State \Return $\mathsf{f}_e^{N_t}, \mathsf{f}_i^{N_t}$
\end{algorithmic}
\end{algorithm}

\subsection{Finite difference Scheme}\label{sec:finite_diff}

To solve the Poisson system~\eqref{eq:poisson} for $\phi$ and compute the electric field $E = -\partial_{z}\phi$ we implement a 4th-order finite difference scheme with homogeneous Dirichlet boundary conditions (i.e., $\phi_{0} = \phi_{N-1} = 0$). For the interior points, the second derivative is approximated as:
\[
    \partial_{zz}\phi(t,z_{i}) \approx \frac{-\phi_{i-2} + 16\phi_{i-1} - 30\phi_{i} + 16\phi_{i+1} -\phi_{i+2}}{12\Delta z^{2}}, \quad \text{for } i=2,3,\dots,N-3\,.
\]
For the near-boundary points ($i=1$ and $i=N-2$), we use the corresponding one-sided 4th-order schemes:
\begin{align*}
    \partial_{zz}\phi(t,z_{1}) &\approx \frac{11\phi_{0} -20\phi_{1} + 6\phi_{2} + 4\phi_{3} - \phi_{4}}{12\Delta z^{2}}\,, \\
    \partial_{zz}\phi(t,z_{N-2}) &\approx \frac{-\phi_{N-5} + 4\phi_{N-4} + 6\phi_{N-3} - 20\phi_{N-2} + 11\phi_{N-1}}{12\Delta z^{2}}\,.
\end{align*}
Similarly, we construct the 4th-order central difference scheme for the first derivative:
\[
    \partial_{z}\phi(t,z_{i}) \approx \frac{\phi_{i-2} - 8\phi_{i-1} + 8\phi_{i+1} - \phi_{i+2}}{12\Delta z}\quad \text{for } i=2,3,\dots,N-3\,,
\]
with the near-boundary approximations:
\begin{align*}
    \partial_{z}\phi(t,z_{1}) &\approx \frac{-3\phi_{0} - 10\phi_{1} + 18\phi_{2} - 6\phi_{3} + \phi_{4}}{12\Delta z}\,, \\
    \partial_{z}\phi(t,z_{N-2}) &\approx \frac{-\phi_{N-5} + 6\phi_{N-4} - 18\phi_{N-3} + 10\phi_{N-2} + 3\phi_{N-1}}{12\Delta z}\,.
\end{align*}

\section{Architecture of the MLP}\label{sec:MLP_architecture}

\begin{figure}[!htb]
    \centering
    \includegraphics[width=1.0\linewidth]{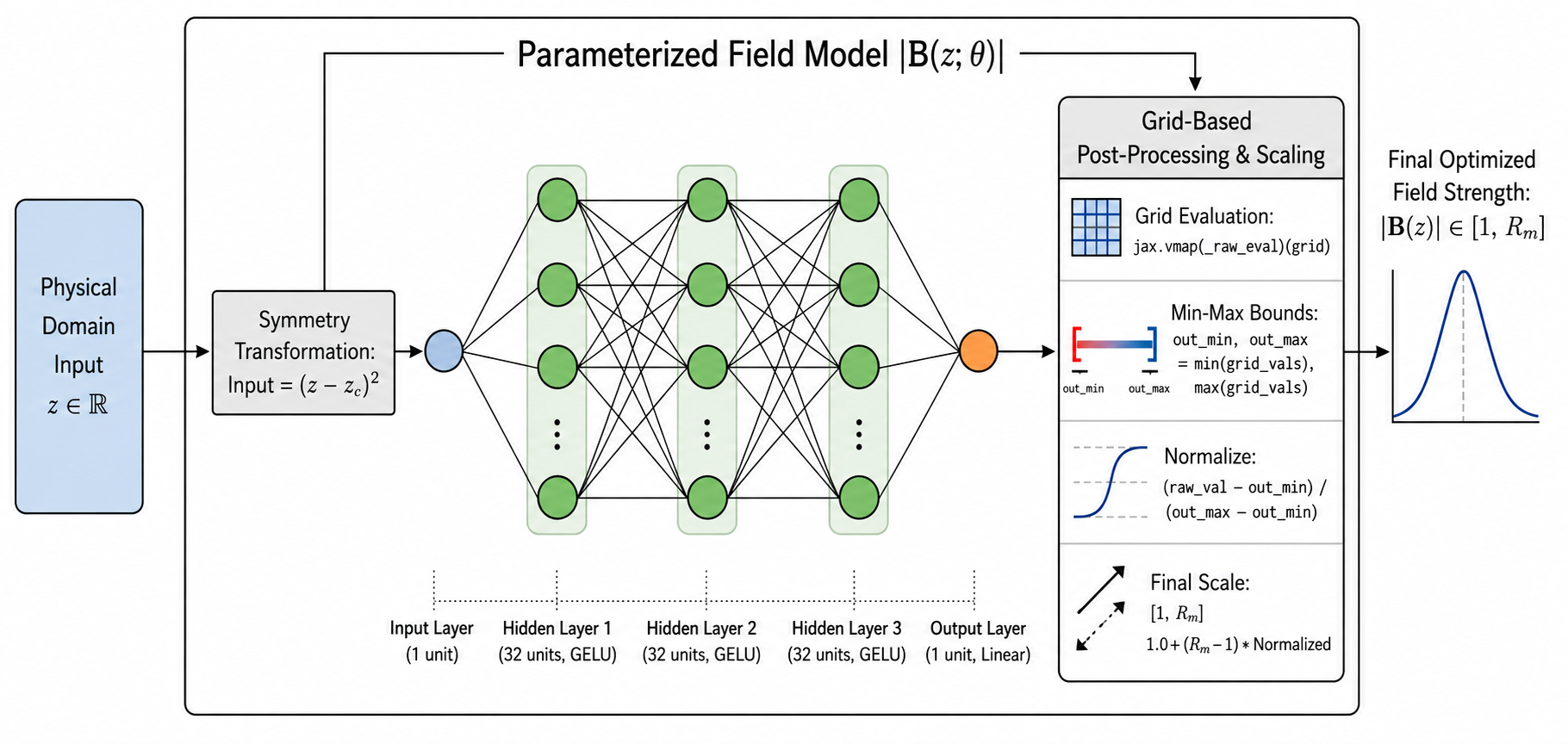}
    \caption{Architecture of the MLP used to define $|\mathbf{B}(z;\theta)|$.}
    \label{fig:MLP_architecture}
\end{figure}

We summarize the architecture of the MLP used to parametrize the magnetic field $|\mathbf{B}|$ in Figure~\ref{fig:MLP_architecture}.

\section{Robustness of the optimization routine}\label{sec:robustness_opt}

We run Algorithm~\ref{alg:optimization_loop} for $100$ different initializations and plot the mean of the different optimal magnetic profiles and a $95\%$ confidence interval for both, the single-species and multi-species regimes.

\begin{figure}[!htb]
    \centering
    \includegraphics[width=0.45\linewidth]{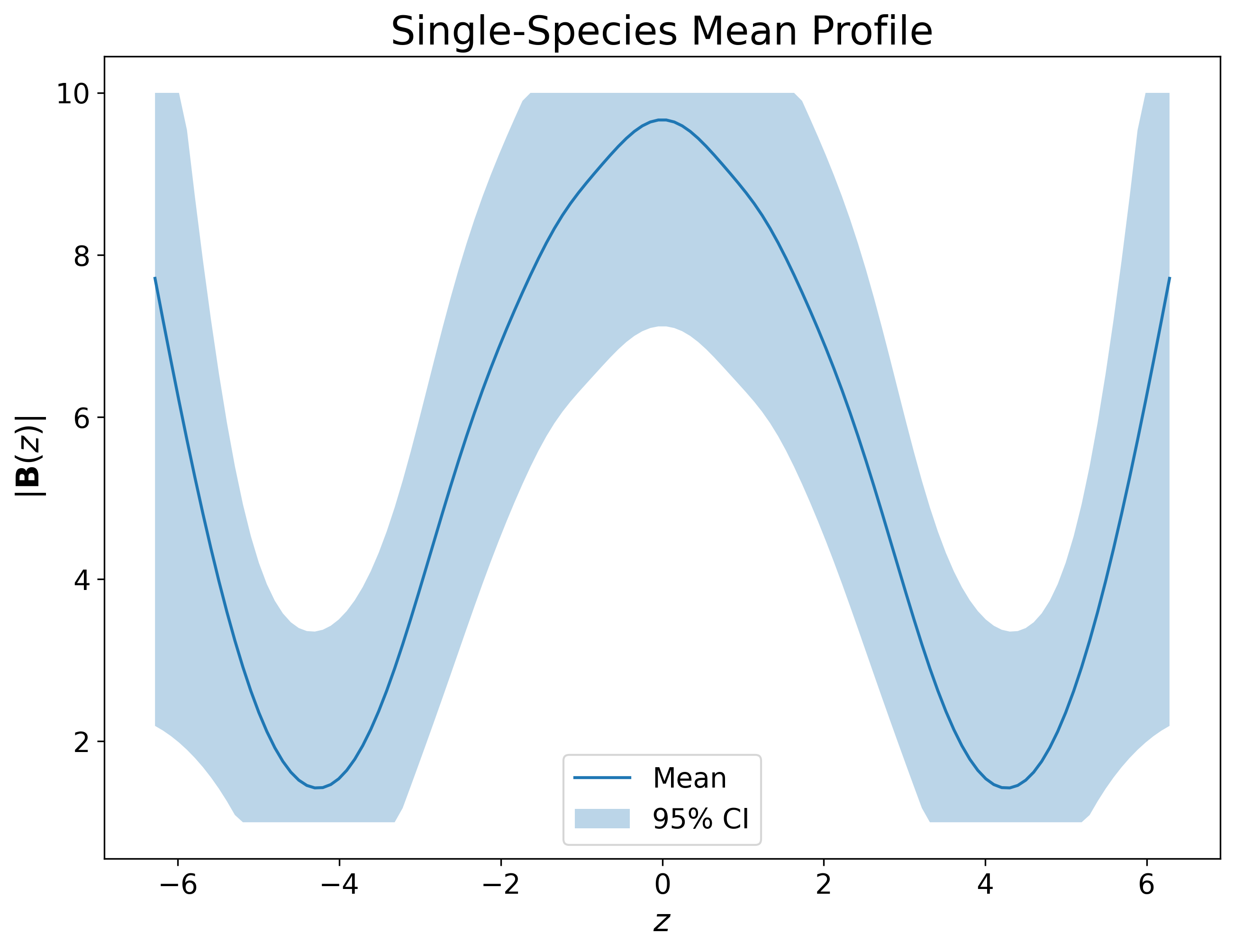}
    \includegraphics[width=0.45\linewidth]{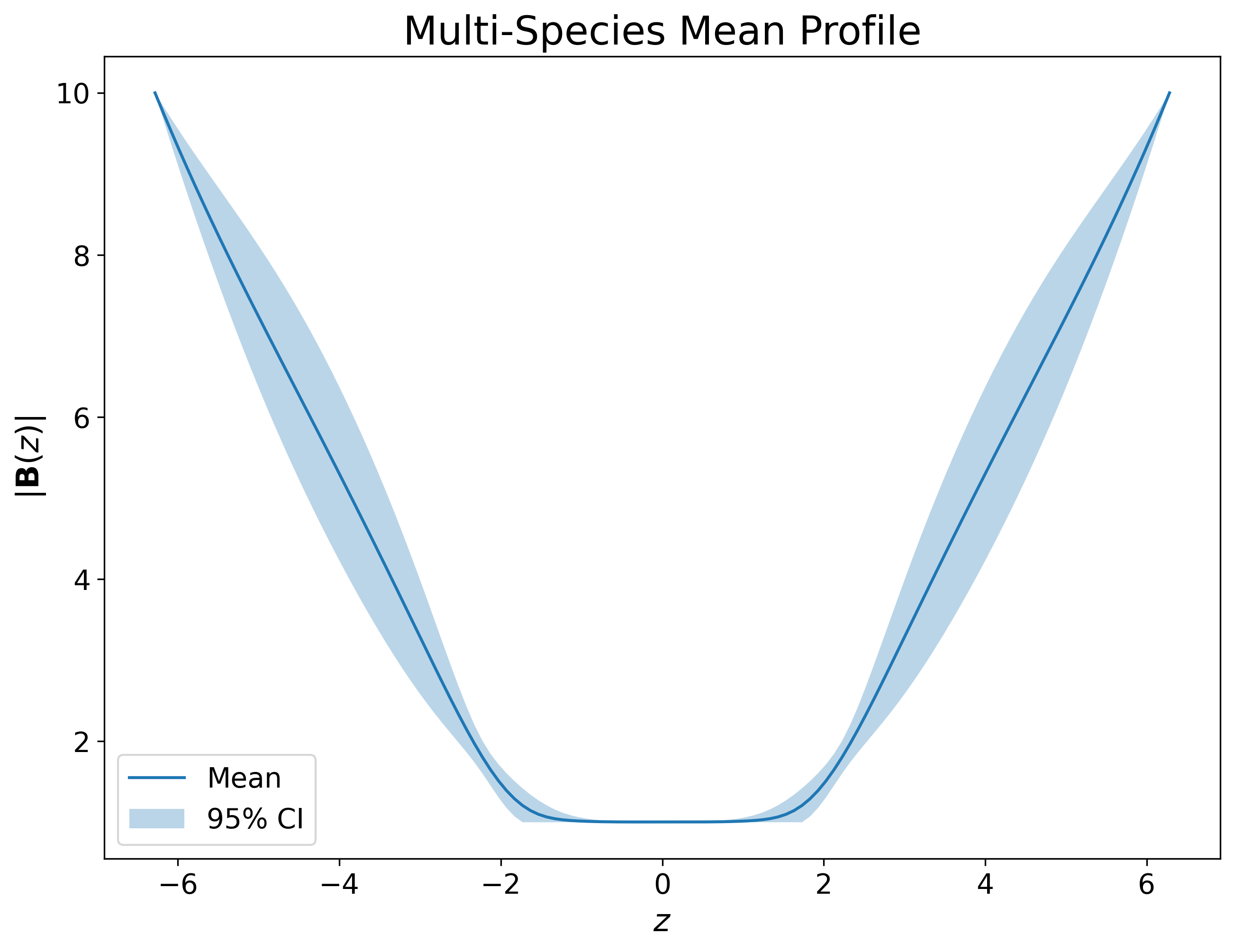}
    \caption{Mean of optimal magnetic profiles $|\mathbf{B}(z;\theta^{\ast})|$ for different initializations $\theta_0$ and shaded area representing a $95\%$ confidence interval. From left to right: single-species and multi-species.}
    \label{fig:heatmap}
\end{figure}

\end{document}